\documentclass[%showpacs, showkeys,
12pt,
preprint,preprintnumbers,nofootinbib,
groupedaddress,superscriptaddress,amsmath,amssymb]{revtex4}
\usepackage{graphicx}% Include figure files
\usepackage{dcolumn}% Align table columns on decimal point
\usepackage{bm}% bold math
\usepackage{amssymb}
\usepackage{amsmath}
\usepackage{epsfig}    
\usepackage{color}
\usepackage{hhline}
\if0
\usepackage{color}
\usepackage[dvips]{graphicx}
\fi
\def\vol#1#2#3{{\bf {#1}} ({#2}) {#3}}
\def\NP{Nucl.~Phys. }
\def\PL{Phys.~Lett. }
\def\PR{Phys.~Rev. }
\def\PRP{Phys.~Rep. }
\def\PRL{Phys.~Rev.~Lett. }

\def\IJMP{Int.~J.~Mod.~Phys. }

\def\EPJ{Eur.~Phys.~J.}

\def\PRD#1#2#3#4{Phys.~Rev.~D~{\bf {#1}}~{(#2)}~no.{#3},~{#4}}

\def\no{\nonumber}

\def\2tvec#1#2{
\left(
\begin{array}{c}
#1  \\
#2  \\   
\end{array}
\right)}

\def\mat2#1#2#3#4{
\left(
\begin{array}{cc}
#1 & #2 \\
#3 & #4 \\
\end{array}
\right)
}

\def\Mat3#1#2#3#4#5#6#7#8#9{
\left(
\begin{array}{ccc}
#1 & #2 & #3 \\
#4 & #5 & #6 \\
#7 & #8 & #9 \\
\end{array}
\right)
}

\def\3tvec#1#2#3{
\left(
\begin{array}{c}
#1  \\
#2  \\   
#3  \\
\end{array}
\right)}

\def\4tvec#1#2#3#4{
\left(
\begin{array}{c}
#1  \\
#2  \\   
#3  \\
#4  \\
\end{array}
\right)}

\def\5tvec#1#2#3#4#5{
\left(
\begin{array}{c}
#1  \\
#2  \\   
#3  \\
#4  \\
#5  \\
\end{array}
\right)}

\def\dq#1{``#1,"}

\def\L{\left}
\def\R{\right}

\def\hbar{\hspace{1mm}\bar{}\hspace{-1mm}h}

\def\eqn#1{
\begin{eqnarray}
#1
\end{eqnarray}
}

\begin{document}
 {\begin{flushright}{APCTP Pre2020 - 032}\end{flushright}}
%%%
  \title{Lepton Anomalous Magnetic Moments in  an $S_4$ Flavor-Symmetric Extra U(1) Model}
  %%%
  \author{Yasuhiro Daikoku}
\email{yasu_daikoku@yahoo.co.jp}
\affiliation{Institute for Theoretical Physics, Kanazawa University, Kanazawa
  920-1192, Japan}

\author{Hiroshi Okada}
\email{hiroshi.okada@apctp.org}
\affiliation{Asia Pacific Center for Theoretical Physics (APCTP) - Headquarters San 31, Hyoja-dong,
Nam-gu, Pohang 790-784, Korea}
\affiliation{Department of Physics, Pohang University of Science and Technology, Pohang 37673, Republic of Korea}

 \if0 
\author{Yasuhiro Daikoku\footnote{E-mail:yasu\_daikoku@yahoo.co.jp}, 
\quad Hiroshi Okada\footnote{E-mail: hiroshi.okada@apctp.org}
\\
%%%
  {\em Institute for Theoretical Physics, Kanazawa University, Kanazawa
  920-1192, Japan.}$^*$ \\
 {\em Asia Pacific Center for Theoretical Physics, Pohang 37673, Republic of Korea}$^\dagger$\\
  {\em Department of Physics, Pohang University of Science and Technology, Pohang 37673, Republic of Korea}$^{\dagger}$
 }
\fi

  \date{\today}
%%%
%  \tableofcontents

\begin{abstract}
We study supersymmetric extra U(1) model with $S_4$ flavor symmetry.
The flavor symmetry not only stabilizes proton but also
suppresses  the  flavor changing processes without raising the supersymmetry breaking scale.
After the flavor symmetry is broken,
the Yukawa hierarchy is realized by the Froggatt-Nielsen mechanism.
The relevant Peccei-Quinn scale for  axion dark matter:
$\frac{f_a}{M_P}\sim 10^{-5}$ accounts for small up quark mass.
The muon mass scale: $\frac{m_\mu}{M_W}\sim 10^{-3}$ is related to 
the $O(10^{-6})$ mass degeneracy of right-handed neutrinos, from 
which we can identify the relevant scale of right-handed neutrino mass for baryon asymmetry of the Universe
as TeV.
Due to the existence of the extra higgsinos,
the discrepancies of  the anomalous magnetic moments of
the muon and the electron between the standard model predictions
and the observations
are explained by the chargino-sneutrino contributions simultaneously.
\end{abstract}
  \maketitle
\newpage

%%%%%%%%%%%%%%%%%%%%%%%%%%%%%%%%%%%%%%%%%%%%%%%
% sec1
%%%%%%%%%%%%%%%%%%%%%%%%%%%%%%%%%%%%%%%%%%%%%%%

\vspace{1cm}

\section{Introduction}

The standard model (SM) is a successful theory of gauge interactions; however,
there are many unsolved issues, such as how to generate the Yukawa hierarchy,
the tiny mass of neutrino and the baryon asymmetry of the Universe (BAU), 
how to stabilize the electroweak scale ($M_W\sim 10^2\mbox{GeV}$) against the
Planck scale ($M_P\sim 10^{18}\mbox{GeV}$) quantum corrections,
why the strong interaction conserves CP  and what dark matter is.
With introducing heavy right-handed neutrino (RHN),
the smallness of neutrino mass and the origin of baryon asymmetry are
accounted for simultaneously by the seesaw mechanism \cite{seesaw} 
and leptogenesis \cite{leptogenesis} respectively.
The well known solution of the Yukawa hierarchy problem is 
the Froggatt-Nielsen mechanism \cite{FN}.
The strong-CP problem is solved by the Peccei-Quinn mechanism \cite{PQ} 
which accommodates a candidate for dark matter; axion \cite{axionDM}.

The elegant solution of the large scale hierarchy problem is supersymmetry (SUSY)
\cite{SUSY} which is the main target of the LHC. The existence of  a light Higgs boson
(such as $125\mbox{GeV}$ \cite{Higgs}) supports the idea of SUSY.
Furthermore, it is well known that the long-standing problem of
the discrepancy of muon anomalous magnetic moment $(g-2)_\mu (\equiv 2a_\mu)$
between the SM prediction and the experimental value~\cite{muon-g-2-exp}:
\eqn{
\Delta a_\mu=a_\mu(\mbox{exp})-a_\mu(\mbox{SM})=(27.9\pm 7.6)\times 10^{-10},
}
can be explained by SUSY \cite{muon-g2SUSY}.
If this explanation is true, then some light sparticles should exist, 
hence SUSY is verifiable  for the LHC or a future collider.

In the minimal supersymmetric standard model (MSSM), as the Higgs
superfields $H^U$ and $H^D$ are vector-like under the SM gauge symmetry
$G_{\mbox{SM}}=\mbox{SU}(3)_c\times \mbox{SU}(2)_W\times \mbox{U}(1)_Y$,
we can introduce a $\mu$ term:
\eqn{
\mu H^UH^D,
}
into the superpotential. The natural size of the parameter $\mu$ is
$O(M_P)$, however, $\mu$ must be $O(M_W)$ in order for the
electroweak gauge symmetry to break. This is  the so-called $\mu$ problem,
which is solved by making the Higgs superfields chiral under a new $\mbox{U}(1)$ symmetry. 
Based on the $E_6$-inspired extra $\mbox{U}(1)$ model \cite{E6},
we can eliminate a fundamental $\mu$ term from the superpotential 
and introduce a trilinear term:
\eqn{
\lambda SH^UH^D,
}
which is converted into an effective $\mu$ term when the singlet $S$
develops an $O(1\mbox{TeV})$ vacuum expectation value (VEV) \cite{mu-prob-u1}.
To prevent the squared masses of sleptons from receiving large D-term contribution of  the $\mbox{U}(1)$, 
we introduce the $\mbox{U}(1)$ under which the lepton doublet $L$ is neutral. 
The new gauge symmetry requires additional Higgs superfields
which contribute to the lepton anomalous magnetic moments.

Though we can account for $(g-2)_\mu$ in this framework,
some problems remain.
The extra  $\mbox{U}(1)$ symmetry requires colored Higgs superfilelds $G,G^c$ to
cancel gauge anomaly
and replaces the baryon- and lepton-number violating
terms in the MSSM  by single G interactions:
\eqn{
GQQ+G^cU^cD^c+GU^cE^c+G^cQL+GN^cD^c,
}
which induce too rapid proton decay.
Successful leptogenesis requires $10^{11}\mbox{GeV}$ scale
RHN, which contradicts with the low reheating temperature as
$T_{RH}<10^7\mbox{GeV}$ which is required to avoid an overproduction of the gravitino
\cite{gravitino-prob}.
Moreover, such a large mass scale is not testable for a collider.
The light slepton induces
too large flavor changing processes such as $\mu\to e+\gamma$
when the soft SUSY breaking squared masses are non-universal as is naively expected
\cite{SUSYFCNC}.
The extra Higgs bosons induce additional contributions to flavor changing process
\cite{HiggsFCNC}.
These problems give additional information about the flavor structure.

To solve these problems, we introduce a flavor symmetry.
There are many candidates of continuous flavor symmetries such as
U(1) \cite{FCNC-flavor-u1} SU(2) \cite{FCNC-flavor-su2}, SU(3)\cite{FCNC-flavor-su3}. 
In this paper, we adopt a discrete $S_4$ flavor symmetry \cite{s4u1, review}.
Assigning the RHNs to singlet and doublet, 
the resonant leptogenesis is realized, which resolves
the contradiction between the reheating temperature and the RHN mass scale
by reducing later \cite{r-leptogenesis}. 
Assigning the left-handed leptons and quarks to singlets and doublets respectively,
the flavor violating processes are suppressed by
a degeneracy of the sfermion mass \cite{s4u1vol2}\cite{s4u1vol3}.
The Yukawa hierarchies of the quarks and the charged leptons are
realized by assigning the right handed fermions to singlets. 
In this case, the discrepancy of the size of representations between left-handed
and right-handed fermions causes suppression of mass
in the same manner as $\mbox{SU}(2)_W$.
We should keep in mind that the hierarchy between fermion mass scale
and the Planck scale is generated by
the discrepancy of the size of the representations of $\mbox{SU}(2)_W$
between the left-handed and right-handed fermions.
We adopt this manner for suppressing single G-interactions, too.
Assigning $G,G^c$ to triplets, the single G interactions are forbidden.
The existence of $S_4$ triplets compels all fermions to consist of three generations
to cancel gauge anomaly,
which is the possible answer to the question, why three generations exist,
(or to the Rabi's question, ``Who ordered that?''). 
The number of generations is fixed to the size of the flavor representation of $G,G^c$ 
which we call ``G Higgs'' (generation-number-imprinted-colored-Higgs).
If the G Higgs decays dominantly to the RHN, 
the seesaw mechanism and leptogenesis may be verified directly at a collider experiment.
The G Higgs is a Rosetta stone to decipher the undeciphered part of flavor physics.
The LHC or a future collider may reveal  what is imprinted in the particle.

Recent observation of the fine structure constant \cite{fine} gives new prediction to
the electron $(g-2)_e$, which has small discrepancy with experimental value~\cite{e-g-2-exp}:
\eqn{
\Delta a_e=a_e(\mbox{exp})-a_e(\mbox{SM})=-(8.7\pm 3.6)\times 10^{-13}.
}
Several papers studying both $g-2$ anomalies based on SUSY
can be found in the literature \cite{both-g-2}.
As the universal new physics contribution gives wrong prediction:
\eqn{
\frac{m^2_\mu a_e(\mbox{NP})}{m^2_ea_\mu(\mbox{NP})}\simeq 1,
} 
for the experimental value:
\eqn{
\frac{m^2_\mu\Delta a_e}{m^2_e\Delta a_\mu}\simeq-14,
}
there should be a non-trivial flavor structure. 
Several attempts to explain it based on flavor symmetry or on other frameworks  
can be found in the literatures (\cite {g-2-flavor} \cite {other-g-2}). 
In this paper we explain both $g-2$ anomalies  simultaneously based on our model.

%%%%%%%%%%%%%%%%%%%%%%%%%%%%%%%%%%%%%%%%%%%%%%%
% sec2
%%%%%%%%%%%%%%%%%%%%%%%%%%%%%%%%%%%%%%%%%%%%%%%
\section{Symmetry breaking}

\subsection{Gauge symmetry}

\begin{table}[htbp]
\begin{center}
\begin{tabular}{|c|c|c|c|c|c|c|c|c|c|c|c|c|c|}
\hline
                      &$Q$    &$U^c$      &$E^c$  &$D^c$     &$L$    &$N^c$ 
                      &$H^D$ &$G^c$      &$H^U$ &$G$        &$S$    &$\Phi$ &$\Phi^c$\\
\hline
$\mbox{SU}(3)_c$             &$3$    &$\bar{3}$  &$1$      &$\bar{3}$ &$1$    &$1$    
                          &$1$     &$\bar{3}$  &$1$      &$3$       &$1$     &$1$   &$1$\\
\hline
$\mbox{SU}(2)_W$            &$2$    &$1$         &$1$      &$1$       &$2$       &$1$
                          &$2$     &$1$        &$2$      &$1$       &$1$      &$1$    &$1$  \\
\hline
$y=6Y$                &$1$    &$-4$       &$6$      &$2$        &$-3$   &$0$   
                          &$-3$   &$2$        &$3$    &$-2$       &$0$      &$0$    &$0$  \\
\hline
$6\sqrt{2/5}Q_\psi$&$1$    &$1$         &$1$      &$1$       &$1$    &$1$    
                          &$-2$   &$-2$        &$-2$   &$-2$      &$4$    &$-1$   &$1$  \\
\hline
$2\sqrt{6}Q_\chi$  &$-1$   &$-1$        &$-1$   &$3$       &$3$    &$-5$    
                          &$-2$   &$-2$        &$2$    &$2$       &$0$    &$5$      &$-5$  \\
\hline
$x=2\sqrt{6}X$      &$1$    &$1$         &$1$      &$2$       &$2$    &$0$    
                          &$-3$   &$-3$       &$-2$   &$-2$      &$5$    &$0$   &$0$  \\
\hline
$z=6\sqrt{2/5}Z$   &$-1$   &$-1$        &$-1$   &$2$       &$2$    &$-4$    
                          &$-1$   &$-1$        &$2$    &$2$        &$-1$  &$4$   &$-4$  \\
\hline
$q_S=(12/\sqrt{10})Q_S$&$1$   &$-1$        &$3$   &$2$       &$0$    &$0$    
                          &$-3$   &$-1$        &$0$    &$-2$        &$3$  &$0$   &$0$  \\
\hline
$Z^\Phi_2$           &$+$    &$+$          &$+$    &$+$        &$+$    &$+$ 
                          &$+$   &$+$          &$+$     &$+$        &$+$    &$-$    &$-$ \\
\hline
\end{tabular}
\end{center}
\caption{$G_{32111}$ assignment of superfields. Here the $x$, $y$, $z$ 
and $q_S$ are charges of
$\mbox{U}(1)_X$, $\mbox{U}(1)_Y$, $\mbox{U}(1)_Z$ 
and $\mbox{U}(1)_S$, and $Y$ is hypercharge. 
The charges of $\mbox{U}(1)_\psi$ and $\mbox{U}(1)_\chi$ ($Q_\psi$ and $Q_\chi$) are also given.}
\end{table}

We extend the gauge symmetry from $G_{\mbox{SM}}$ to
$G_{32111}=G_{\mbox{SM}}\times \mbox{U}(1)_S\times \mbox{U}(1)_Z$, and
introduce new superfields $N^c,S,G,G^c$ which are
embedded in the {\bf 27} representation of $E_6$ with
quark and lepton superfields $Q,U^c,D^c,L,E^c$ and Higgs superfields
$H^U,H^D$.
Here, $N^c$ is  the RHN, $S$ is the $G_{\mbox{SM}}$ singlet, and $G,G^c$ are colored Higgses.   
The two $\mbox{U}(1)$ charges,  $X$ and $Z$ are defined by the
 linear combinations of $Q_\psi$ and $Q_\chi$  as
\eqn{
X=\frac{\sqrt{15}}{4}Q_\psi+\frac14Q_\chi, \quad
Z=-\frac14Q_\psi+\frac{\sqrt{15}}{4}Q_\chi,
}
where
$E_6 \supset \mbox{SO}(10) \times \mbox{U}(1)_\psi \supset \mbox{SU}(5)\times
\mbox{U}(1)_\chi \times \mbox{U}(1)_\psi$.

The definition of $\mbox{U}(1)_S$ charge  under
which the left-handed lepton is neutral is given by
\eqn{
Q_S=\sqrt{\frac25}Y+\sqrt{\frac35}X,
}
which is automatically anomaly free. 
Note that  it is impossible to embed $G_{32111}$ in $E_6$.
The charge assignment of the superfields are given in Table 1.
To break $\mbox{U}(1)_Z$, we add new vector-like superfields $\Phi$,  $\Phi^c$,
where $\Phi^c$ is the same representation as the RHN $N^c$ under the $G_{32111}$,
and its anti-representation $\Phi$ originates from $\bf{27}^*$.
To discriminate between $N^c$ and $\Phi^c$, we introduce $Z^\Phi_2$ symmetry under which
$\Phi^c$ and $\Phi$ are odd.
The invariant superpotential under these symmetries is given by
\eqn{
W_{32111}&=&W_0+W_S+W_G+W_\Phi , \\
W_0&=&Y^UH^UQU^c+Y^DH^DQD^c+Y^EH^DLE^c+Y^NH^U_LN^c
+\frac{1}{M_P}Y^M\Phi\Phi N^cN^c , \\
W_S&=&kSGG^c+\lambda SH^UH^D , \\
W_G&=&Y^{QQ}GQQ+Y^{UD}G^cU^cD^c+Y^{UE}GU^cE^c
+Y^{QL}G^cQL+Y^{DN}GD^cN^c , \\
W_\Phi&=&M_\Phi\Phi\Phi^c +\frac{1}{M_P}(\Phi\Phi^c)^2+\cdots ,
}
where $M_P=2.4353\times 10^{18} \mbox{GeV}$ is the reduced Planck scale.
Since the interactions in $W_S$ drive the squared mass of $S$ to be negative through
renormalization group equations, 
the $\mbox{U}(1)_S$ symmetry is broken spontaneously
and the $\mbox{U}(1)_S$ gauge boson $Z'$ acquires the mass
\eqn{
m(Z')\simeq \frac{\sqrt{5}}{2} g_S\L<S\R> ,
}  
where $\L<H^D\R>\ll \L<S\R>$ is assumed based on the experimental
constraint for $Z'_\psi$~\cite{PDG}: 
\eqn{
m(Z'_\psi)> 3900\mbox{GeV}.
}
The constraint for the $Z'$ mass is not far from this bound.
In this paper we assume $\L<H^{U,D}\R>/\L<S\R>\sim O(10^{-2})$.

If $M_\Phi=0$ in $W_\Phi$ and the origin of the potential 
$V(\Phi,\Phi^c)$ is unstable, then $\Phi,\Phi^c$ develop large VEVs
along the D-flat direction of $\L<\Phi\R>=\L<\Phi^c\R>=V$,
$\mbox{U}(1)_Z$ is broken, and the $\mbox{U}(1)_Z$ gauge boson $Z''$
acquires the mass
\eqn{
m(Z'')=\frac43\sqrt{\frac52}g_ZV.
}
After the gauge symmetry breaking, since the R-symmetry defined by
\eqn{
R=Z^\Phi_2 \exp\L[\frac{i\pi}{4}(q_S-2y+3z)\R],
}
remains unbroken, the lightest SUSY particle (LSP) is stable.

\subsection{Flavor symmetry}

\begin{table}[htbp]
\begin{center}
\begin{tabular}{|c|c|c|c|c|c|c|c|c|c|c|c|c|c|c|c|}
\hline
           &$Q_i$     &$Q_3$  &$U^c_1$  &$U^c_2$   &$U^c_3$  &$D^c_1$  &$D^c_2$ 
           &$D^c_3$ &$L_i$   &$L_3$      &$E^c_1$   &$E^c_2$  &$E^c_3$  &$N^c_i$ 
           &$N^c_3$ \\
\hline
$S_4$   &$2$       &$1$      &$1$        &$1'$        &$1$        &$1'$        &$1$
           &$1$       &$2$      &$1$        &$1'$        &$1$        &$1$        &$1$ 
            &$1$ \\
\hline
$Z^Q_2$ &$-$       &$+$      &$+$        &$-$         &$+$        &$-$        &$+$
            &$+$       &$+$      &$+$        &$+$         &$+$        &$+$        &$+$ 
            &$+$ \\
\hline
$Z^a_7$ &$1$       &$0$      &$0$        &$0$         &$0$        &$0$        &$0$
            &$0$       &$2$      &$2$        &$-1$       &$-1$      &$-2$      &$0$ 
            &$0$ \\
\hline
$Z^b_7$ &$0$       &$0$      &$1$        &$0$         &$0$        &$2$        &$1$
            &$2$       &$1$      &$1$        &$1$        &$-1$      &$1$         &$0$ 
            &$0$ \\
\hline
$Z^c_7$ &$0$       &$0$      &$0$        &$0$         &$0$        &$1$        &$1$
            &$0$       &$0$      &$1$        &$0$        &$0$        &$-1$       &$0$ 
            &$0$ \\
\hline
$Z^d_7$ &$0$       &$0$      &$0$        &$1$         &$0$        &$0$        &$0$
            &$0$       &$1$      &$0$        &$0$        &$0$        &$0$        &$0$ 
            &$0$ \\
\hline
$Z_9$    &$0$       &$0$      &$1$        &$0$         &$0$        &$0$        &$0$
            &$0$       &$0$      &$0$        &$0$        &$0$        &$0$        &$0$ 
            &$0$ \\
\hline
$Z_5$    &$0$       &$0$      &$1$        &$0$         &$0$        &$0$        &$0$
            &$0$       &$0$      &$0$        &$0$        &$0$        &$0$        &$0$ 
            &$0$ \\
\hline
$Z^F_2$ &$+$       &$+$      &$+$        &$+$         &$+$        &$+$        &$+$
            &$+$       &$+$      &$+$        &$+$        &$+$        &$+$        &$+$ 
            &$+$ \\
\hline
\end{tabular}
\begin{tabular}{|c|c|c|c|c|c|c|c|c|c|c|c|c|}
\hline
            &$S_1$ &$S_2$     &$S_3$     &$H^U_i$    &$H^U_3$    &$H^D_i$     &$H^D_3$  
            &$G_a$    &$G^c_a$ &$\Phi_i$    &$\Phi_3$    &$\Phi^c_a$    \\
\hline
$S_4$    &$1$   &$1$       &$1$        &$2$         &$1$           &$2$           &$1$
            &$3$       &$3$        &$2$         &$1$           &$3$           \\
\hline
$Z^Q_2$ &$+$  &$+$       &$+$        &$+$         &$+$           &$+$           &$+$
            &$+$       &$+$        &$+$         &$+$           &$+$           \\
\hline
$Z^a_7$ &$2$  &$0$       &$0$        &$1$         &$0$            &$-1$         &$0$
            &$0$       &$0$        &$0$         &$0$           &$0$           \\
\hline
$Z^b_7$ &$0$  &$0$       &$0$        &$0$         &$0$            &$0$           &$0$
            &$0$       &$0$        &$-2$       &$-2$         &$0$           \\
\hline
$Z^c_7$ &$0$  &$0$       &$0$        &$0$         &$0$            &$0$           &$0$
            &$0$       &$0$        &$0$         &$0$           &$0$           \\
\hline
$Z^d_7$ &$2$  &$0$       &$0$        &$1$         &$0$            &$-1$         &$0$
            &$0$       &$0$        &$0$         &$0$           &$0$           \\
\hline
$Z_9$    &$0$  &$0$       &$0$        &$0$         &$0$            &$0$          &$0$
            &$0$       &$0$        &$0$         &$0$           &$0$           \\
\hline
$Z_5$    &$0$  &$0$       &$0$        &$0$         &$0$            &$0$          &$0$
            &$0$       &$0$        &$0$         &$0$           &$0$           \\
\hline
$Z^F_2$ &$+$  &$+$       &$+$        &$+$         &$+$           &$+$           &$+$
            &$+$       &$+$        &$+$         &$+$           &$+$           \\
\hline
\end{tabular}
\begin{tabular}{|c|c|c|c|c|c|c|c|c|c|}
\hline
           &$X$          &$F^A_i$  &$F^A_3$   &$F^B_i$      &$F^B_3$    &$F^C_i$  
           &$F^C_3$   &$R_i$     &$T_i$          \\
\hline
$S_4$   &$1$          &$2$        &$1$         &$2$           &$1$           &$2$
           &$1$          &$2$        &$2$            \\
\hline
$Z^Q_2$ &$-$        &$+$        &$+$          &$+$           &$+$           &$+$
            &$+$        &$+$        &$+$              \\
\hline
$Z^a_7$ &$0$        &$-1$       &$-1$         &$-1$         &$-1$           &$-1$
            &$-1$      &$-1$       &$-3$              \\
\hline
$Z^b_7$ &$-1$      &$0$        &$0$           &$-1$         &$-1$           &$0$
            &$0$       &$0$        &$0$              \\
\hline
$Z^c_7$ &$0$       &$-1$       &$-1$         &$0$           &$0$             &$0$
            &$0$       &$0$        &$0$              \\
\hline
$Z^d_7$ &$0$       &$0$        &$0$          &$0$           &$0$             &$-1$
            &$-1$       &$0$        &$0$              \\
\hline
$Z_9$    &$0$       &$0$        &$0$          &$0$           &$0$             &$0$
            &$0$       &$-1$        &$-5$              \\
\hline
$Z_5$    &$0$       &$0$        &$0$          &$0$           &$0$             &$0$
            &$0$       &$-1$        &$-1$              \\
\hline
$Z^F_2$ &$+$        &$+$        &$-$          &$+$           &$-$           &$+$
            &$-$        &$+$        &$+$              \\
\hline
\end{tabular}
\end{center}
\caption{$S_4\times Z^Q_2\times Z^a_7\times Z^b_7\times Z^c_7\times Z^d_7\times Z_9
\times Z_5\times Z^F_2$ assignment of superfields. Here the index $i$ of the $S_4$
doublets runs $i=1,2$,  and the index $a$ of the $S_4$ triplets runs $a=1,2,3$.
The details of $S_4$ are given in Ref. \cite{s4}.}
\end{table}

The superpotential defined in Eq.(10)-(14) has the following problems.
As the interaction $W_G$ induces a proton decay that is too fast,
it must be strongly suppressed. The mass parameter $M_\Phi$ in $W_\Phi$
must be forbidden in order to allow for $\mbox{U}(1)_Z$ symmetry breaking.
In $W_0$, the contributions to flavor-changing processes from the extra
Higgs bosons must be suppressed.
These problems should be solved by flavor symmetry.

If we introduce $S_4$ flavor symmetry and assign $G,G^c$ to be triplets,
then $W_G$ [defined in Eq. (13)] is forbidden.
This is because any products of doublets and singlets of $S_4$ do not contain triplets. 
However, as the existence of  a G Higgs with a lifetime longer than $0.1s$ spoils the
success of big bang nucleosynthesis (BBN), the $S_4$ symmetry must be broken.
Therefore we assign $\Phi^c$ to be a triplet 
in order to break $\mbox{U}(1)_Z$ and $S_4$ at the same time
and assign $\Phi$ to be a doublet and a singlet  in order to forbid $M_\Phi\Phi\Phi^c$.   
With these assignments, $S_4$ symmetry is broken due to the VEV of
$\Phi^c$ and the effective trilinear terms which correspond to $W_G$ are induced from nonrenormalizable terms.
The size of the VEV of $\Phi$ is fixed by the superpotential
\eqn{
W_{X\Phi}=\frac{1}{M^{11}_P}[X^{14}+Y^\Phi X^{10}(\Phi\Phi^c)^2
+X^{8}(\Phi\Phi^c)^3+X^{6}(\Phi\Phi^c)^4+X^{4}(\Phi\Phi^c)^5
+X^{2}(\Phi\Phi^c)^6+(\Phi\Phi^c)^7],\no\\
}
and the soft SUSY-breaking terms as follows:
\eqn{
\frac{\L<X\R>}{M_P}=\frac{\L<\Phi\R>}{M_P}=10^{-\frac54},
}
where $X$ is a gauge singlet.
We assume that the global minimum of the potential $V(X,\Phi,\Phi^c)$
is at the $S_3$-symmetric vacuum and along with the D-flat direction of $\mbox{U}(1)_Z$
as follows:
\eqn{
\L<\Phi_1\R>=\L<\Phi_2\R>=0, \quad 
\L<X\R>=\L<\Phi_3\R>=\sqrt{3}\L<\Phi^c_1\R>
=\sqrt{3}\L<\Phi^c_2\R>=\sqrt{3}\L<\Phi^c_3\R>=V\equiv 10^{-\frac54}M_P.
}

The assignments of the other superfields are determined based on the following
criteria: 1) the mass matrices of quarks and leptons are consistent with the observed
mass hierarchies and the Cabibbo-Kobayashi-Maskawa (CKM) 
and Maki-Nakagawa-Sakata (MNS) matrices;
2) the third-generation Higgses $H^U_3,H^D_3$ are specified as 
the dominant component of MSSM Higgses;
3) the experimental constraints for the flavor violating processes are satisfied;
4) the resonant leptogenesis mechanism works;
5) an accidental Peccei-Quinn U(1) global symmetry is included;
6) two anomalies in $(g-2)_e$ and $(g-2)_\mu$ are accounted for.
The representation of all superfields under
the flavor symmetry is given in Table 2.

In order to realize the Yukawa hierarchies, we introduce gauge singlet
flavon superfields $F^{A,B,C}_{i,3},R_i,T_i$ and fix their VEVs as follows:
\eqn{
&&\frac{\L<F^A_{1,2,3}\R>}{M_P}=\epsilon^3(c_a,s_a,1), \quad
\frac{\L<F^B_{1,2,3}\R>}{M_P}=\epsilon^3(\alpha c_b,\beta s_b,1), \quad
\frac{\L<F^C_{1,2,3}\R>}{M_P}=\epsilon^3(c_c,s_c,1), \no \\
&&\frac{\L<R_i\R>}{M_P}=\epsilon^5(c_R,s_R), \quad
\frac{\L<T_i\R>}{M_P}=\epsilon^5(c_T,s_T), \quad
\epsilon=0.1, \quad |\alpha|=|\beta|=1, \no \\
&&c_x\equiv\cos\theta_x, \quad s_x\equiv\sin\theta_x \quad (x=a,b,c,R,T,\cdots ) ,
}
where $\alpha$ and $\beta$ are complex.
In this paper, we assume that the original Lagrangian has CP
symmetry and all parameters in it are real.
Therefore the complex VEVs given in Eq. (22) induce
spontaneous CP violation.

The superpotentials of $F^X_{i,3}$ are given by
\eqn{
W_{F^X}=\frac{1}{M^4}\L[(E^X_2)^2+(F^X_3)^2E^X_2+(F^X_3)^4\R]E^X_3,
}
for $X=A,B,C$, respectively, where $E^X_2=(F^X_1)^2+(F^X_2)^2$ and
$E^X_3=3(F^X_1)^2F^X_2-(F^X_2)^3$ are $S_4$ invariants.
The $S_4$ invariants $E^2_2E_3, F^2_3E_2E_3,F^4_3E_3$ 
have nine different spurions $a_{1,2,\cdots,9}$:
\eqn{
E^2_2E_3&=& a_1F^6_1F_2+a_2F^4_1F^3_2+a_3F^2_1F^5_2+a_4F^7_2, \\
F^2_3E_2E_3&=&a_5F^4_1F_2F^2_3+a_6F^2_1F^3_2F^2_3+a_7F^5_2F^2_3, \\
F^4_3E_3&=&a_8F^2_1F_2F^4_3+a_9F^3_2F^4_3,
}
where $a_1=3,a_2=5,a_3=1,a_4=-1,a_5=3,a_6=2,a_7=-1,a_8=3,a_9=-1$ 
and $F_i$ is the $S_4$ doublet and $F_3$ is the singlet.
Since the number of spurions ($=9$) is larger than the dimension ($=3$) of
the vector space spanned by the nine charge vectors as follows:
\eqn{
&&a_1:(6,1,0), \quad a_2:(4,3,0), \quad a_3:(2,5,0), \quad a_4:(0,7,0), \quad
a_5:(4,1,2), \quad  a_6: (2,3,2),  \\
&&a_7:(0,5,2), \quad a_8:(2,1,4), \quad a_9:(0,3,4); \quad 
\mbox{for} \quad F_1(-1,0,0), \quad F_2(0,-1,0), \quad F_3(0,0,-1),\no
}
spontaneous CP violation is not forbidden (see \cite{SCPV}).

The scale of VEV is fixed as follows. If the superpotential of the gauge singlet superfield $\Psi$ is
given by
\eqn{
W=\frac{\Psi^n}{nM^{n-3}_P},
}
then the potential of $\Psi$ is given by
\eqn{
V(\Psi)=m^2_\Psi|\Psi|^2-\L(\frac{A\Psi^n}{M^{n-3}_P}+h.c.\R)+\frac{|\Psi^{n-1}|^2}{M^{2n-6}_P},
}
where $m_\Psi \sim A\sim m_{\mbox{SUSY}}\sim O(1)\mbox{TeV}$ is assumed.
At the global minimum $\L<\Psi\R>\neq 0$, because each of the terms in
the potential should be balanced, the scale of the VEV is fixed by
\eqn{
\frac{\L<\Psi\R>}{M_P}=\L(\frac{m_{\mbox{SUSY}}}{M_P}\R)^{1/(n-2)}.
}

We assume that the effect of SUSY breaking in the hidden sector is
mediated by gravity and induces soft SUSY-breaking terms in the observable 
sector. Since these terms are non-universal in general,
large flavor-changing processes are induced by the sfermion exchange.
From the experimental constraints on them, the assignments of
quarks and leptons under the flavor symmetry are restrictive.

After the flavor symmetry breaking, the soft breaking scalar squared mass matrices
become non-diagonal.  For the Higgs scalars, this gives the mixing mass terms
\eqn{
V &\supset& m^2_{UB}\epsilon^3(H^U_3)^*(c_cH^U_1+s_cH^U_2)
+m^2_{DB}\epsilon^3(H^D_3)^*H^D_i(c_cH^D_1+s_cH^D_2)\no\\
&+& m^2\epsilon^5(S_1)^*(c_SS_2+s_SS_3)+h.c. ,
} 
which compel the extra Higgs scalars to develop VEVs as
\eqn{
\L<H^U_i\R>=N_U\epsilon^3(c_c,s_c)v_u, \quad
\L<H^D_i\R>=N_D\epsilon^3(c_c,s_c)v_d, \quad
\L<S_1\R>=O(\epsilon^5)v_s,
}
where we put
\eqn{
&&\L<H^U_3\R>=v_u=150.7\mbox{GeV}, \quad
\L<H^D_3\R>=v_d=87.0\mbox{GeV}, \quad
\sqrt{v^2_u+v^2_d}=v=174.0\mbox{GeV}, \no \\
&&\L<S_2\R>=c_sv_s, \quad \L<S_3\R>=s_sv_s, \quad
\L<S\R>=\sqrt{\L<S_2\R>^2+\L<S_3\R>^2}=v_s \geq 9581\mbox{GeV}.
}
The constraint for $v_s$ is derived from Eq.(16) and the assumption:
$g_S(1\mbox{TeV})=g_Y(1\mbox{TeV})=0.3641$.
As the same effects affect the flavons, the VEV directions given in Eq.(21)
are perturbed as follows:
\eqn{
\L<\Phi_1\R>\sim \L<\Phi_2\R>\sim O(\epsilon^6)V , \quad
\sqrt{3}\L<\Phi^c_a\R>=(1+O(\epsilon^6))V, \quad
\L<F^A_1\R>=\epsilon^3(c_a+O(\epsilon^6))M_P, \quad \cdots
}
and so on. 
Note that the dominant parts of the scalar squared mass matrices
of the extra Higgs and G Higgs are diagonal and degenerated.
Due to the smallness of VEVs of the extra Higgs bosons, the superpartners
of the extra Higgs and G Higgs also have diagonal and degenerated mass matrices.
Therefore the trace of $S_4$ flavor symmetry is imprinted in
their mass spectra which may be testable for the LHC or a future collider.

\subsection{Accidental Peccie-Quinn symmetry}

In this paper, we adopt axion-flavon unification scenario \cite{AF} which
can be embedded into the non-abelian flavor symmetric models, for example:
$A_4$ \cite{AF-A4}, $D_6$ \cite{AF-D6},  $T'$ \cite{AF-Tprime}, $SL_2(F_3)$ \cite{AF-SL}
and into the supersymmetric model: \cite{AF-SUSY} too.
As the cut off scale of our model is the Planck scale $M_P$,
the way to embed the axion into flavon is restrictive.
For the allowed region of the PQ scale:
$10^8 (\mbox{SN1987A})<f_a<10^{12} (\mbox{Dark Mater})\mbox{GeV}$,
the order of the effective Yukawa coupling is given by
\eqn{
Y_{\mbox{eff}}=\frac{f_a}{M_P}\sim 10^{-10}\sim 10^{-6},
}
which is not sizable for the second and third generation Yukawa couplings.
In this paper we identify $Y_{\mbox{eff}}$ as the up quark Yukawa coupling \cite{up-AF}.

At the leading order, the potential of flavons $R_i,T_i$ is given by
\eqn{
V=m^2_R|R|^2+m^2_T|T|^2-\frac{A}{M^2_P}(R^4T+h.c.)
+\frac{1}{M^4_P}[|R^4|^2+|4R^3T|^2],
}
where $m_R\sim m_T\sim m_{\mbox{SUSY}}$ is assumed 
and  unimportant $S_4$ indexes  are omitted. 
As this potential is invariant under the redefinition of fields as
\eqn{
R\to e^{i\theta}R, \quad T\to e^{-4i\theta}T, \quad 
U^c_1\to e^{-i\theta}U^c_1,
}
it has an accidental Peccei-Quinn  $\mbox{U}(1)$ symmetry
\cite{axion-quality},
hence the strong-CP problem is solved by the Peccei-Quinn mechanism.
We assume that the F-flat direction:
$R=0,T=\infty$ is stabilized due to the positive squared mass $m^2_T$.
The phase of flavon $a$ defined by
\eqn{
R=f_ae^{ia/f_a}, \quad T=f_ae^{-4ia/f_a}, \quad f_a=10^{13}\mbox{GeV},
}
takes the role of axion, where $f_a$ is a $U(1)_{PQ}$ breaking scale.

Taking account of next leading order superpotential
\eqn{
W_{PQB}=\frac{(RT^4)^3(F^B)^3X^4}{M^{19}_P}
=\frac{\epsilon^{13}}{M^{12}_P}(RT^4)^3,
}
this accidental $\mbox{U}(1)$ symmetry is explicitly broken
and the low energy axion potential is modified as
\eqn{
V_a=-\Lambda^4_{\mbox{QCD}}\cos\L(\theta_0+\frac{a}{f_a}\R)
-\frac{\epsilon^{13}m_{\mbox{SUSY}}f^{15}_a}{M^{12}_P}\cos(45a/f_a),
}
which sifts the global minimum from $\theta_{\mbox{QCD}}=0$ to
\eqn{
\theta_{\mbox{QCD}}=\frac{45\epsilon^{13}m_{\mbox{SUSY}}f^{15}_a}
{M^{12}_P\Lambda_{\mbox{QCD}}}\sim 10^{-27}.
}
As the experimental constraint on the neutron electric moment:
$\theta_{\mbox{QCD}}<10^{-10}$ is satisfied,
this $\mbox{U}(1)$ symmetry has sufficient quality. 

The density parameter of the coherent oscillation of the axion is
evaluated as
\eqn{
\frac{\Omega_a}{\Omega_{\mbox{CDM}}}
\sim \theta^2_i\L(\frac{f_a}{10^{12}\mbox{GeV}}\R)^{\frac76},
}
where $\theta_i$ is the initial value of the strong-CP phase
when the QCD potential of the axion is switched on.
In this paper, we assume $\theta_i\sim 0.3$ and dark matter is dominated by the axion.
As the domain wall number of this model is $N_{DW}=1$, our model is free from
a domain wall problem of the axion.
Furthermore, we assume that the flavor symmetry is not recovered both
during inflation (e.g. due to the negative Hubble induced mass terms)  
and after reheating (due to the low reheating temperature)
hence a domain wall problem of the discrete flavor symmetry is avoided \cite{domain-wall}.  
As the flavon multiplets, including the axino which is the superpartner of the axion, 
have TeV scale mass,
the thermal production of them is suppressed due to the low reheating temperature.

%%%%%%%%%%%%%%%%%%%%%%%%%%%%%%%%%%%%%%%%%%%%%%%
% sec3
%%%%%%%%%%%%%%%%%%%%%%%%%%%%%%%%%%%%%%%%%%%%%%%
\section{Quark sector}

The superpotential of the quark sector is given by
\eqn{
W=H^U_3QY^UU^c+H^D_3QY^DD^c
+H^U_iQY^{UI}_iU^c+H^D_iQY^{DI}_iD^c,
}
where the Yukawa matrices are given by
\eqn{
Y^U&=&
\Mat3{\epsilon^6Y^U_1c_R}{-\epsilon^3Y^U_2s_c}{\epsilon^{16}Y^U_4\alpha c_b}
{\epsilon^6Y^U_1s_R}{\epsilon^3Y^U_2c_c}{\epsilon^{16}Y^U_4\beta s_b}
{\epsilon^{25}Y^U_5}{\epsilon^{22}Y^U_6}{Y^U_3}, \quad
Y^{UI}_i=
\Mat3{\epsilon^9Y^{UI}_{i11}}{\epsilon^6Y^{UI}_{i12}}{\epsilon^{19}Y^{UI}_{i13}}
{\epsilon^9Y^{UI}_{i21}}{\epsilon^6Y^{UI}_{i22}}{\epsilon^{19}Y^{UI}_{i23}}
{\epsilon^{28}Y^{UI}_{i31}}{\epsilon^{25}Y^{UI}_{i32}}{\epsilon^3Y^{UI}_{i33}}, \\
Y^D&=&\Mat3{\epsilon^5Y^D_1s_a}{\epsilon^4Y^D_2c_a}{\epsilon^4Y^D_4\alpha c_b}
{-\epsilon^5Y^D_1c_a}{\epsilon^4Y^D_2s_a}{\epsilon^4Y^D_4\beta s_b}
{\epsilon^{24}Y^D_5}{\epsilon^{23}Y^D_6}{\epsilon^2Y^D_3}, \quad
Y^{DI}_i=\Mat3{\epsilon^{23}Y^{DI}_{i11}}{\epsilon^{22}Y^{DI}_{i12}}{\epsilon^{22}Y^{DI}_{i13}}
{\epsilon^{23}Y^{DI}_{i21}}{\epsilon^{22}Y^{DI}_{i22}}{\epsilon^{22}Y^{DI}_{i23}}
{\epsilon^{42}Y^{DI}_{i31}}{\epsilon^{41}Y^{DI}_{i32}}{\epsilon^{20}Y^{DI}_{i33}}.
}
As the K\"ahler potential receives the effect of the flavor violation, the superfields
must be redefined as
\eqn{
U^c&\to& V_K(U)U^c, \quad D^c\to V_K(D)D^c, \quad Q\to V_K(Q)Q, \\
V_K(U)&=&\Mat3{1}{\epsilon^{9}k^U_{12}s_{R-c}}{\epsilon^{10}k^U_{13}\alpha^U_1}
{\epsilon^{9}k^U_{12}s_{R-c}}{1}{\epsilon^{7}k^U_{23}\alpha^U_2}
{\epsilon^{10}k^U_{13}(\alpha^U_1)^*}{\epsilon^{7}k^U_{23}(\alpha^U_2)^*}{1}, \\
V_K(D)&=&\Mat3{1}{i\epsilon^7k^D_{12}s_{2b}}{\epsilon^7k^D_{13}\alpha^D_1}
{-i\epsilon^7k^D_{12}s_{2b}}{1}{\epsilon^6k^D_{23}\alpha^D_2}
{\epsilon^7k^D_{13}(\alpha^D_2)^*}{\epsilon^6k^D_{23}(\alpha^D_2)^*}{1}, \\
V_K(Q)&=&\Mat3{1}{\epsilon^6k^Q_{12}\alpha_Q}{\epsilon^4k^Q_3c_b\alpha}
{\epsilon^6k^Q_{12}(\alpha_Q)^*}{1}{\epsilon^4k^Q_3s_b\beta}
{\epsilon^4k^Q_3c_b\alpha^*}{\epsilon^4k^Q_3s_b\beta^*}{1},
}
in order to get canonical kinetic terms \cite{Kahler}.
As the result, the quark Yukawa matrices are redefined as
\eqn{
(Y^U)'&=&V^T_K(Q)Y^UV_K(U), \quad (Y^{UI})'=V^T_K(Q)Y^{UI}V_K(U), \no \\
(Y^D)'&=&V^T_K(Q)Y^DV_K(D), \quad (Y^{DI})'=V^T_K(Q)Y^{DI}V_K(D), \\
(Y^U)'&=&\Mat3{Y^U_1c_R\epsilon^6}{-Y^U_2s_c\epsilon^3}{O(\epsilon^{4})}
{Y^U_1s_R\epsilon^6}{Y^U_2c_c\epsilon^3}{O(\epsilon^{4})}
{O(\epsilon^{10})}{O(\epsilon^{7})}{Y^U_3}, \quad
(Y^{UI}_i)' =\Mat3{\epsilon^{9}}{\epsilon^{6}}{\epsilon^{7}}
{\epsilon^{9}}{\epsilon^{6}}{\epsilon^{7}}
{\epsilon^{13}}{\epsilon^{10}}{\epsilon^{3}} , \\
(Y^D)'&=&\Mat3{Y^D_1s_a\epsilon^5}{Y^D_2c_a\epsilon^4}{Y^D_4\alpha c_b\epsilon^4}
{-Y^D_1c_a\epsilon^5}{Y^D_2s_a\epsilon^4}{Y^D_4\beta s_b\epsilon^4}
{O(\epsilon^{9})}{O(\epsilon^{8})}{Y^D_3\epsilon^2}, \quad
(Y^{DI}_i)' =\Mat3{\epsilon^{23}}{\epsilon^{22}}{\epsilon^{22}}
{\epsilon^{23}}{\epsilon^{22}}{\epsilon^{22}}
{\epsilon^{27}}{\epsilon^{26}}{\epsilon^{20}}.
}
Likewise, the Higgs superfields must be redefined as
\eqn{ 
H^U&\to& V_K(H^U)H^U, \quad H^D\to V_K(H^D)H^D, \\
V_K(H^U)&=&\Mat3{1}{\rho_1\epsilon^6}{k_Uc_c\epsilon^3}
{\rho^*_1\epsilon^6}{1}{k_Us_c\epsilon^3}
{k_Uc_c\epsilon^3}{k_Us_c\epsilon^3}{1}, \\
V_K(H^D)&=&\Mat3{1}{\rho_2\epsilon^6}{k_Dc_c\epsilon^3}
{\rho^*_2\epsilon^6}{1}{k_Ds_c\epsilon^3}
{k_Dc_c\epsilon^3}{k_Ds_c\epsilon^3}{1},
}
hence the order of the elements in $(Y^{DI})'$ is modified to
\eqn{
(Y^{DI}_i)'' =(Y^{DI}_i)'+O(\epsilon^3)(Y^D)' 
=\Mat3{\epsilon^{8}}{\epsilon^{7}}{\epsilon^{7}}
{\epsilon^{8}}{\epsilon^{7}}{\epsilon^{7}}
{\epsilon^{12}}{\epsilon^{11}}{\epsilon^{5}}.
}
On the other hand, the changes of $(Y^U)',(Y^D)',(Y^{UI}_i)'$ are negligible. 
Since the contributions to the quark mass matrices from
the extra Higgs bosons through $(Y^{UI}_i)' ,(Y^{DI}_i)' $ are negligible,
the quark mass matrices are approximated as
\eqn{
M'_U=(Y^U)'v_u, \quad M'_D=(Y^D)'v_d.
}
These matrices are diagonalized by the superfields redefinitions
\eqn{
U&\to& L_UU, \quad D\to L_DD, \quad U^c\to R_UU^c, \quad D^c\to R_DD^c,} 
\eqn{
(L_U)^T&=&\Mat3{c_c}{s_c}{\epsilon^4}
{-s_c}{c_c}{\epsilon^4}
{\epsilon^4}{\epsilon^4}{1}, \\
R_U&=&\Mat3{1}{(Y^U_1/Y^U_2)s_U\epsilon^3}{\epsilon^{10}}
{-(Y^U_1/Y^U_2)s_U\epsilon^3}{1}{\epsilon^{7}}
{\epsilon^{10}}{\epsilon^{7}}{1} ,  \quad \theta_U=\theta_R-\theta_c,}
% \\
\eqn{(L_D)^T&=&\Mat3{s_a}{-c_a}{-(Y^D_4/Y^D_3)(\alpha c_bs_a-\beta s_bc_a)\epsilon^2}
{c_a}{s_a}{-(Y^D_4/Y^D_3)(\alpha c_bc_a+\beta s_bs_a)\epsilon^2}
{(Y^D_4/Y^D_3)\alpha^*c_b\epsilon^2}
{(Y^D_4/Y^D_3)\beta^*s_b\epsilon^2}{1}, \\
R_D&=&\Mat3{1}{\epsilon^5}{\epsilon^5}
{\epsilon^5}{1}{\epsilon^4}
{\epsilon^5}{\epsilon^4}{1},
}
from which we get
\eqn{
&&L^T_UM'_UR_U=\mbox{diag}(m_u,m_c,m_t)
=\mbox{diag}(Y^U_1c_U\epsilon^6,Y^U_2\epsilon^3,Y^U_3)v_u, \\
&&L^T_DM'_DR_D=\mbox{diag}(m_d,m_s,m_b)
=\mbox{diag}(Y^D_1\epsilon^5,Y^D_2\epsilon^4,Y^D_3\epsilon^2)v_d, \\
&&V_{CKM}=L^\dagger_U L_D
=\Mat3{s_{a-c}}{c_{a-c}}{(c_cc_b\alpha^*+s_cs_b\beta^*)r_D\epsilon^2}
{-c_{a-c}}{s_{a-c}}{-(s_cc_b\alpha^*-c_cs_b\beta^*)r_D\epsilon^2}
{-(s_ac_b\alpha-c_as_b\beta)r_D\epsilon^2}
{-(c_ac_b\alpha+s_as_b\beta)r_D\epsilon^2}{1}, \no \\
&&s_{a-c}=\sin(\theta_a-\theta_c), \quad r_D=Y^D_4/Y^D_3.
}
The experimental values of the CKM matrix elements: 
\eqn{
&&\Mat3{|V_{ud}|}{|V_{us}|}{|V_{ub}|}
{|V_{cd}|}{|V_{cs}|}{|V_{cb}|}
{|V_{td}|}{|V_{ts}|}{|V_{tb}|}
=\Mat3{0.974}{0.227}{0.361\times 10^{-2}}
{0.226}{0.973}{4.05\times 10^{-2}}
{0.854\times 10^{-2}}{3.98\times 10^{-2}}{1} , \\
&&J=\mbox{Im}(V_{us}V_{cb}V^*_{ub}V^*_{cs})=3.00\times 10^{-5},
}
are realized by tuning the five parameters:
$\theta_{a,b,c},r_D,\mbox{arg}(\alpha\beta^*)$.
The experimental values of quark running masses at $1$ TeV \cite{PDG}\cite{quark-mass}:
\eqn{
&&m_u=1.17\times 10^{-3}, \quad m_c=0.543, \quad m_t=148.1, \\
&&m_d=2.40\times 10^{-3}, \quad m_s=4.9\times 10^{-2}, \quad m_b=2.41 
\quad (\mbox{GeV}), 
}
are realized by the Planck scale boundary values:
\eqn{
&&|Y^U_1c_U|=1.5,  \quad |Y^U_2|=0.71, \quad |Y^U_3|=0.28,  \no \\
&&|Y^D_1|=0.38,  \quad |Y^D_2|=0.78, \quad |Y^D_3|=0.38,
}
where we have used the renormalization factors given in Ref. \cite{s4u1vol2}.
As these parameters are consistent with the assumption that
all of the factor $Y$ are $O(1)$, the quark mass hierarchy is realized without fine-tuning.
In our definition, $Y=O(1)$ means that $10^{-0.5}<Y<10^{0.5}$ is satisfied.
The soft SUSY-breaking squared mass matrices of the squarks are given by
\eqn{
\frac{m^2_U}{m^2}&=&
\Mat3{O(1)}{\epsilon^9}{\epsilon^{10}}
{\epsilon^9}{O(1)}{\epsilon^7}
{\epsilon^{10}}{\epsilon^7}{O(1)}, \\
\frac{m^2_D}{m^2}&=&
\Mat3{O(1)}{\epsilon^7}{\epsilon^7}
{\epsilon^7}{O(1)}{\epsilon^6}
{\epsilon^7}{\epsilon^6}{O(1)}, \\
\frac{m^2_Q}{m^2}&=&
\Mat3{1}{\epsilon^6}{\epsilon^4}
{\epsilon^6}{1}{\epsilon^4}
{\epsilon^4}{\epsilon^4}{O(1)},
}
and the squark A-term matrices are given by
\eqn{
V&\supset& -v_uUA_UU^c-v_dDA_DD^c+h.c. \\
A_U&=&\Mat3{A^U_1c_R\epsilon^6}{-A^U_2s_c\epsilon^3}{A^U_4 O(\epsilon^{4})}
{A^U_1s_R\epsilon^6}{A^U_2c_c\epsilon^3}{A^U_4 O(\epsilon^{4})}
{A^U_5O(\epsilon^{10})}{A^U_6O(\epsilon^7)}{A^U_3}, \\
A_D&=&\Mat3{A^D_1s_a\epsilon^5}{A^D_2c_a\epsilon^4}{A^D_4\alpha c_b\epsilon^4}
{-A^D_1c_a\epsilon^5}{A^D_2s_a\epsilon^4}{A^D_4\beta s_b\epsilon^4}
{A^D_5O(\epsilon^9)}{A^D_6O(\epsilon^8)}{A^D_3\epsilon^2},
}
where these matrices are defined for canonically normalized superfields.
The sizes of parameters $m,A^X_n$ are assumed to be  $O(\mbox{TeV})$. 
After the diagonalization of the Yukawa matrices, the squared mass and A-term matrices are
given by 
\eqn{
(m^2_U)_{SCKM}&=&R^\dagger_Um^2_UR_U
=m^2\Mat3{O(1)}{\epsilon^3}{\epsilon^{10}}
{\epsilon^3}{O(1)}{\epsilon^7}
{\epsilon^{10}}{\epsilon^7}{O(1)}, \\
(m^2_D)_{SCKM}&=&R^\dagger_Dm^2_DR_D
=m^2\Mat3{O(1)}{\epsilon^5}{\epsilon^5}
{\epsilon^5}{O(1)}{\epsilon^4}
{\epsilon^5}{\epsilon^4}{O(1)} ,\\
(m^2_Q)_{SCKM}&=&L^\dagger_{(U,D)}m^2_QL_{(U,D)}
=m^2\Mat3{1}{\epsilon^{(6,4)}}{\epsilon^{(4,2)}}
{\epsilon^{(6,4)}}{1}{\epsilon^{(4,2)}}
{\epsilon^{(4,2)}}{\epsilon^{(4,2)}}{O(1)} , \\
(A_U)_{SCKM}&=&L^T_UA_UR_U
=\Mat3{A^U_1c_U\epsilon^6}{A\epsilon^{9}}{A\epsilon^{4}}
{A\epsilon^6}{A^U_2\epsilon^3}{A\epsilon^{4}}
{A\epsilon^{10}}{A\epsilon^{7}}{A^U_3} , \\
(A_D)_{SCKM}&=&L^T_DA_DR_D
=\Mat3{A^D_1\epsilon^5}{A\epsilon^{8}}{A\epsilon^{4}}
{A\epsilon^{9}}{A^D_2\epsilon^4}{A\epsilon^{4}}
{A\epsilon^{7}}{A\epsilon^{6}}{A^D_3\epsilon^2}.
}
The off-diagonal elements of the squark mass matrices contribute to
the flavor and CP violation through the squark exchange, on which severe constraints
are imposed. With the mass insertion approximation,
the most stringent bound for the squark mass $M_Q$ is given by $\epsilon_K$ as
\eqn{
\sqrt{\frac{\mbox{Im}[(m^2_Q)_{12}(m^2_D)_{12}]}{M^4_Q}}
=\epsilon^{4.5}<4.4\times 10^{-4}\L(\frac{M_Q}{\mbox{TeV}}\R)\quad \to \quad
M_Q>72\mbox{GeV},
}
where $M_Q=M(\mbox{gluino})=M(\mbox{squark})$ is assumed \cite{SUSYFCNC}.
This bound is very weak and the SUSY flavor-changing-neutral-current (FCNC)
problem is solved.
The contribution to the FCNC from the extra Higgs is also suppressed enough due to the
small $Y^{UI},Y^{DI}$.
For example, the constraint from $D^0-\bar{D}^0$ mixing 
gives very weak constraint: $m_H>2.5\mbox{GeV}$ for the extra Higgs boson mass\cite{HiggsFCNC}.

%%%%%%%%%%%%%%%%%%%%%%%%%%%%%%%%%%%%%%%%%%%%%%%
% sec4
%%%%%%%%%%%%%%%%%%%%%%%%%%%%%%%%%%%%%%%%%%%%%%%

\section{Lepton sector}

The superpotential of the lepton is given by
\eqn{
W=H^U_3LY^NN^c+H^D_3LY^EE^c+H^D_iLY^{EI}_iE^c
+M_NN^cY^MN^c.
}
After the redefinition of superfields:
\eqn{
E^c\to V_K(E)E^c, \quad N^c\to V_K(N)N^c, \quad L\to V_K(L)L,\quad
H^U\to V_K(H^U)H^U, \quad H^D\to V_K(H^D)H^D,\no\\
}
the kinetic terms are canonically normalized and
the Yukawa matrices are given by
\eqn{
(Y^N)'&=&\epsilon^6
\Mat3{\alpha c_bc_cY^N_1+\beta s_bs_cY^N_2+Y^N_5}
{\alpha c_bs_cY^N_4+\beta s_bc_c Y^N_3}{Y^N_7(\alpha c_bs_c+\beta s_bc_c)}
{\alpha c_bs_cY^N_3+\beta s_bc_cY^N_4}
{\alpha c_bc_c Y^N_2+\beta s_bs_c Y^N_1+Y^N_5}{Y^N_7(\alpha c_bc_c-\beta s_bs_c)}
{Y^N_6(\alpha c_bs_a+\beta s_bc_a)}
{Y^N_6(\alpha c_bc_a-\beta s_bs_a)}{Y^N_{8}(\alpha c_bc_a+\beta s_bs_a)+Y^N_{9}}, \no\\\\
(Y^E)'&=&V^T_K(L)Y^EV_K(E)
=\Mat3{\epsilon^5Y^E_1s_c}{\epsilon^3Y^E_2c_c}{\epsilon^{8}}
{-\epsilon^5Y^E_1c_c}{\epsilon^3Y^E_2s_c}{\epsilon^{8}}
{\epsilon^{11}}{\epsilon^{9}}{\epsilon^2Y^E_3}, \\
(Y^{EI}_1)'&=&V^T_K(L)Y^{EI}_1V_K(E)
=\Mat3{\epsilon^8}{Y^{EI}_2}{\epsilon^{11}}
{-\epsilon^2Y^{EI}_1}{\epsilon^{6}}{\epsilon^{11}}
{\epsilon^{8}}{\epsilon^{6}}{\epsilon^{5}}, \\
(Y^{EI}_2)'&=&V^T_K(L)Y^{EI}_2V_K(E)
=\Mat3{\epsilon^2Y^{EI}_1}{\epsilon^{6}}{\epsilon^{11}}
{\epsilon^8}{Y^{EI}_2}{\epsilon^{11}}
{\epsilon^{8}}{\epsilon^{6}}{\epsilon^{5}}, \\
(Y^M)'&=&\Mat3{Y^M_1}{\epsilon^6}{\epsilon^6}
{\epsilon^6}{Y^M_1}{\epsilon^6}
{\epsilon^6}{\epsilon^6}{Y^M_3}.
}
The charged lepton mass matrix is given by
\eqn{
M'_E=\L<H^D_3\R>(Y^E)'+\L<H^D_i\R>(Y^{EI}_i)'
=\Mat3{\epsilon^5(Y^E_1+Y^{EI}_1N_D)s_c}
{\epsilon^3(Y^E_2+Y^{EI}_2N_D)c_c}{\epsilon^{8}}
{-\epsilon^5(Y^E_1+Y^{EI}_1N_D)c_c}{\epsilon^3(Y^E_2+Y^{EI}_2N_D)s_c}{\epsilon^{8}}
{\epsilon^{11}}{\epsilon^{9}}{\epsilon^2Y^E_3}v_d.\no\\
}
This matrix is diagonalized by the superfields redefinition
\eqn{
E&\to&L_E E, \quad E^c\to R_EE^c , \\
(L_E)^T&=&\Mat3{s_c}{-c_c}{\epsilon^{6}}
{c_c}{s_c}{\epsilon^{6}}
{\epsilon^{6}}{\epsilon^{6}}{1}, \\
R_E&=&\Mat3{1}{\epsilon^{14}}{\epsilon^{9}}
{\epsilon^{14}}{1}{\epsilon^{7}}
{\epsilon^{9}}{\epsilon^{7}}{1},
}
from which we get
\eqn{
L^T_EM'_ER_E=
\mbox{diag}(m_e,m_\mu,m_\tau)
=\mbox{diag}\L((Y^E_1+Y^{EI}_1N_D)\epsilon^5,(Y^E_2+Y^{EI}_2N_D)\epsilon^3,Y^E_3\epsilon^2\R)v_d.
}
The experimental values of the charged lepton running masses  at $1$ TeV~\cite{quark-mass}:
\eqn{
m_e=4.895\times 10^{-4}\  (\mbox{GeV}), \quad
m_\mu=0.1033  (\mbox{GeV}), \quad
m_\tau=1.757 \ (\mbox{GeV}),
}
are realized by setting the parameters at $1$ TeV as
\eqn{
Y^E_1+Y^{EI}_1N_D=0.56,  \quad Y^E_2+Y^{EI}_2N_D=1.19, \quad Y^E_3=2.02.
}

The seesaw neutrino mass matrix is given by
\eqn{
M_\nu=(L_E)^T(Y^N)'[M_N(Y^M)']^{-1}[(Y^N)']^TL_E,
}
which is diagonalized as
\eqn{
U^T_{\mbox{MNS}}M_\nu U_{\mbox{MNS}}=\mbox{diag}(m_1,m_2,m_3),
}
where $U_{\mbox{MNS}}$ is the Maki-Nakagawa-Sakata matrix.
Since there are too many parameters in $M_\nu$, we cannot give
any prediction for the neutrino mass and the elements of MNS matrix
except for the mass scales
\eqn{
M_N=M_P\L(\frac{\L<X\R>}{M_P}\R)^{10}\L(\frac{\L<\Phi_3\R>}{M_P}\R)^2=10^{-15}M_P
=1\mbox{TeV}, \quad
m_\nu=\frac{(\epsilon^6 v_u)^2}{M_N}\sim 10^{-2} \mbox{eV}.
}
The soft SUSY-breaking squared-mass matrices of the sleptons are given by  
\eqn{
\frac{m^2_E}{m^2}=
\Mat3{O(1)}{\epsilon^8}{\epsilon^9}
{\epsilon^8}{O(1)}{\epsilon^{11}}
{\epsilon^9}{\epsilon^{11}}{O(1)}, \quad
\frac{m^2_L}{m^2}=
\Mat3{1}{\epsilon^6}{\epsilon^6}
{\epsilon^6}{1}{\epsilon^6}
{\epsilon^6}{\epsilon^6}{O(1)},
}
and the slepton A-term matrices are given by
\eqn{
V&\supset& -\L<H^D_3\R>EA^EE^c-\L<H^D_i\R>EA^{EI}_iE^c+h.c. ,\\
A^E&=&\Mat3{A^E_1s_c\epsilon^5}{A^E_2c_c\epsilon^3}{A\epsilon^8}
{-A^E_1c_c\epsilon^5}{A^E_2s_c\epsilon^3}{A\epsilon^8}
{A\epsilon^{11}}{A\epsilon^{9}}{A^E_3\epsilon^2} , \\
A^{EI}_1&=&\Mat3{A\epsilon^8}{A^{EI}_2}{A\epsilon^{11}}
{-\epsilon^2 A^{EI}_1}{A\epsilon^{6}}{A\epsilon^{11}}
{A\epsilon^{8}}{A\epsilon^{6}}{A\epsilon^{5}} ,\quad
A^{EI}_2=\Mat3{\epsilon^2A^{EI}_1}{A\epsilon^6}{A\epsilon^{11}}
{A\epsilon^8}{A^{EI}_2}{A\epsilon^{11}}
{A\epsilon^{8}}{A\epsilon^{6}}{A\epsilon^{5}},
}
where these matrices are defined for canonically normalized superfields.
We define the effective A-term matrix as follows:
\eqn{
&&(A^E)'v_d=A^E\L<H^D_3\R>+A^{EI}_1\L<H^D_1\R>+A^{EI}_2\L<H^D_2\R>, \\
&&(A^E)'=\Mat3{(A^E_1)'s_c\epsilon^5}{(A^E_2)'c_c\epsilon^3}{A\epsilon^8}
{-(A^E_1)'c_c\epsilon^5}{(A^E_2)'s_c\epsilon^3}{A\epsilon^8}
{A\epsilon^{11}}{A\epsilon^{9}}{A^E_3\epsilon^2}, \no \\
&&(A^E_1)'=A^E_1+N_DA^{EI}_1, \quad
(A^E_2)'=A^E_2+N_DA^{EI}_2.
}
After the diagonalization of the charged lepton Yukawa matrix,
the squared-mass matrices and the effective A-term matrix are given by  
\eqn{
(m^2_E)_{\mbox{SMNS}}&=&R^\dagger_Em^2_ER_E
=m^2\Mat3{O(1)}{\epsilon^8}{\epsilon^9}
{\epsilon^8}{O(1)}{\epsilon^7}
{\epsilon^9}{\epsilon^7}{O(1)} , \\
(m^2_L)_{\mbox{SMNS}}&=&L^\dagger_Em^2_EL_E
=m^2\Mat3{1}{\epsilon^6}{\epsilon^6}
{\epsilon^6}{1}{\epsilon^6}
{\epsilon^6}{\epsilon^6}{O(1)} , \\
(A^E)_{\mbox{SMNS}}&=&L^T_E(A^E)'R_E=
\Mat3{A^E_1\epsilon^5}{A\epsilon^{9}}{A\epsilon^{8}}
{A\epsilon^{11}}{A^E_2\epsilon^3}{A\epsilon^{8}}
{A\epsilon^{11}}{A\epsilon^{9}}{A^E_3\epsilon^2}.
}
As the $(1,1)$ element of $(A^E)_{\mbox{SMNS}}$ is real
at leading order,  the SUSY contribution to the
electric dipole moment of the electron is negligible.
Based on the consideration of the lepton flavor violation, the most
stringent bound for the slepton mass $M_L$ is given by $\mu\to e+\gamma$ as
\eqn{
\frac{v_d}{M_L}\epsilon^9<1.4\times 10^{-6}\L(\frac{M_L}{300\mbox{GeV}}\R)
\sqrt{\frac{(\mbox{Br}(\mu\to e\gamma))_{\mbox{exp}}}{4.2\times 10^{-13}}}
\quad \to \quad M_L>4.3\mbox{GeV},
}
where $M_L=M(\mbox{slepton})=M(\mbox{photino})$ is assumed.

For the canonically normalized superfields, the RHN mass matrix is given by
\eqn{
M_R=M_N(Y^M)' , 
} 
whose eigenvalues $M_1,M_2,M_3$ give the degenerated mass spectrum of RHNs as follows:
\eqn{
M_1\simeq M_2=M_1(1+\epsilon^6)&\to& \delta_N=\frac{M_2-M_1}{M_1}\sim \epsilon^6.
}
In this paper, we assume $N^c_3$ is the heaviest RHN, hence $M_2<M_3$. 
The right-handed sneutrinos have the same spectrum.
In the early Universe, the out-of-equilibrium decay of $n^c_1$ and $N^c_1$
generates a lepton asymmetry which is transformed into a
baryon asymmetry by the electroweak sphaleron process.
Following Ref.\cite{SUSY-leptogenesis}, the baryon asymmetry is given by
\eqn{
B_f\sim -\frac{\kappa \epsilon_{CP}}{3g_*},
}
where $g_*=340$ is the degree of freedom of radiation,
$\kappa$ is the dilution factor given by
\eqn{
\kappa&\sim& \frac{1}{K\ln K}, \\
K&=&\frac{\Gamma(M_1)}{2H(M_1)}, \
\Gamma(M_1)=\frac{K_{11}M_1}{8\pi}, \quad
H(M_1)=\sqrt{\frac{\pi^2g_*M^4_1}{90M^2_P}}, \
K_{ij}=\sum^3_{l=1}(Y^N_{li})^*(Y^N_{lj}),
}
and $\epsilon_{CP}$ is given by
\eqn{
\epsilon_{CP}=-\frac{1}{2\pi}\frac{\mbox{Im}(K^2_{12})}{K_{11}}
\L(\frac{2\sqrt{x}}{x-1}+\sqrt{x}\ln\frac{1+x}{x}\R)\simeq
-\frac{\mbox{Im}(K^2_{12})}{2\pi K_{11}\delta_N}, \quad
x=\frac{M^2_2}{M^2_1}\simeq 1+2\delta_N.
}
From the order estimations
\eqn{
K_{12}\sim K_{11}\sim \epsilon^{12}, \quad
K\sim 6\L(\frac{\mbox{TeV}}{M_1}\R), \quad
\epsilon_{CP}\sim 10^{-6}, \quad M_1\sim 1000\mbox{GeV},
}
we get the observed baryon asymmetry, $B_f\sim 10^{-10}$.
This mechanism works even at the low reheating temperature 
as $T_{RH}<10^7\mbox{GeV}$ which is required for avoiding gravitino
overproduction.

%%%%%%%%%%%%%%%%%%%%%%%%%%%%%%%%%%%%%%%%%%%%%%%
% sec5
%%%%%%%%%%%%%%%%%%%%%%%%%%%%%%%%%%%%%%%%%%%%%%%

\section{Higgs sector}

\subsection{Higgs bosons}

For the canonically normalized superfields,
the superpotential of the Higgses up to $O(\epsilon^3)$-terms is given by
\eqn{
W&=&\lambda_2S_2H^U_3H^D_3+\lambda_3S_3H^U_3H^D_3 \no \\
&+&\lambda_4S_2(H^U_1H^D_1+H^U_2H^D_2)
+\lambda_5S_3(H^U_1H^D_1+H^U_2H^D_2) \no \\
&+&\epsilon^3\lambda_6S_2(c_cH^U_1+s_cH^U_2)H^D_3 
+\epsilon^3\lambda_7S_3(c_cH^U_1+s_cH^U_2)H^D_3 \no \\
&+&\epsilon^3\lambda_8S_2H^U_3(c_cH^D_1+s_cH^D_2) 
+\epsilon^3\lambda_9S_3H^U_3(c_cH^D_1+s_cH^D_2),
}
from which we get the Higgs potential as
\eqn{
V&=&-m^2_{H^U_3}|H^U_3|^2+m^2_{H^D_3}|H^D_3|^2+m^2_{S_1}|S_1|^2
-m^2_{S_2}|S_2|^2-m^2_{S_3}|S_3|^2 \no \\
&+&m^2_{H^U}(|H^U_1|^2+|H^U_2|^2)
+m^2_{H^D}(|H^D_1|^2+|H^D_2|^2)
-m^2_{S_4}(S^*_2S_3+S_2S^*_3) \no \\
&-&\epsilon^3m^2_{BU}[(c_cH^U_1+s_cH^U_2)^*H^U_3+h.c.]
-\epsilon^3m^2_{BD}[(c_cH^D_1+s_cH^D_2)^*H^D_3+h.c.] \no}
%%%
\eqn{
&-&A_2[S_2H^U_3H^D_3+h.c.]-A_3[S_3H^U_3H^D_3+h.c.] \no \\
&-&A_4[S_2(H^U_1H^D_1+H^U_2H^D_2)+h.c.]
-A_5[S_3(H^U_1H^D_1+H^U_2H^D_2)+h.c.] \no \\
&-&\epsilon^3A_6[S_2(c_cH^U_1+s_cH^U_2)H^D_3+h.c.]
-\epsilon^3A_7[S_3(c_cH^U_1+s_cH^U_2)H^D_3] \no \\
&-&\epsilon^3A_8[S_2H^U_3(c_cH^D_1+s_cH^D_2) +h.c.]
-\epsilon^3A_9[S_3H^U_3(c_cH^D_1+s_cH^D_2)+h.c.] \no }
\vspace{-1cm}
%%%
\eqn{
&+&\L|\lambda_2H^U_3H^D_3+\lambda_4(H^U_1H^D_1+H^U_2H^D_2) \R. \no \\
&+&\L. \epsilon^3\lambda_6(c_cH^U_1+s_cH^U_2)H^D_3
+\epsilon^3\lambda_8H^U_3(c_cH^D_1+s_cH^D_2)\R|^2 \no \\
&+&\L|\lambda_3H^U_3H^D_3+\lambda_5(H^U_1H^D_1+H^U_2H^D_2) \R. \no \\
&+&\L. \epsilon^3\lambda_7(c_cH^U_1+s_cH^U_2)H^D_3
+\epsilon^3\lambda_9H^U_3(c_cH^D_1+s_cH^D_2)\R|^2 \no \\
&+&\L|\lambda_2S_2H^D_3+\lambda_3S_3H^D_3
+\epsilon^3\lambda_8S_2(c_cH^D_1+s_cH^D_2)
+\epsilon^3\lambda_9S_3(c_cH^D_1+s_cH^D_2)\R|^2 \no}
%%%
\eqn{&+&\L|\lambda_2S_2H^U_3+\lambda_3S_3H^U_3
+\epsilon^3\lambda_6S_2(c_cH^U_1+s_cH^U_2)
+\epsilon^3\lambda_7S_3(c_cH^U_1+s_cH^U_2)\R|^2 \no \\
&+&\L|\lambda_4S_2H^D_1+\lambda_5S_3H^D_1
+\epsilon^3\lambda_6c_cS_2H^D_3
+\epsilon^3\lambda_7c_cS_3H^D_3\R|^2 \no \\
&+&\L|\lambda_4S_2H^D_2+\lambda_5S_3H^D_2
+\epsilon^3\lambda_6s_cS_2H^D_3
+\epsilon^3\lambda_7s_cS_3H^D_3\R|^2 \no \\
&+&\L|\lambda_4S_2H^U_1+\lambda_5S_3H^U_1
+\epsilon^3\lambda_8c_cS_2H^U_3
+\epsilon^3\lambda_9c_cS_3H^U_3\R|^2 \no \\
&+&\L|\lambda_4S_2H^U_2+\lambda_5S_3H^U_2
+\epsilon^3\lambda_8s_cS_2H^U_3
+\epsilon^3\lambda_9s_cS_3H^U_3\R|^2 \no \\
&+&\frac18 g^2_Y\L[|H^U_a|^2-|H^D_a|^2\R]^2
+\frac18g^2_2\sum^3_{A=1}\L[(H^U_a)^*\sigma_AH^U_a
+(H^D_a)^*\sigma_AH^D_a\R]^2 \no \\
&+&\frac92g^2_s\L[|S_a|^2-|H^D_a|^2\R]^2,
}
where
\eqn{
g_s=\frac{\sqrt{10}}{12}g_S.
}
With the definition of the charged, the CP-even neutral,
and the CP-odd neutral Higgs bosons
\eqn{
H^U_a=\2tvec{u^+_a}{u_a+ip_a}, \quad H^D_a=\2tvec{d_a+iq_a}{d^-_a} , \quad
S_a=s_a+ir_a,
}
their mass matrices are defined as follows:
\eqn{
V&\supset&(u^+_a,d^+_a)
\mat2{m^2_{ab}(U^+U^-)}{m^2_{ab}(U^+D^-)}
{m^2_{ab}(D^+U^-)}{m^2_{ab}(D^+D^-)}
\2tvec{u^-_b}{d^-_b} \no \\
&+&(p_a,q_a,r_a)
\Mat3{m^2_{ab}(UU)}{m^2_{ab}(UD)}{0}
{m^2_{ab}(DU)}{m^2_{ab}(DD)}{0}
{0}{0}{m^2_{ab}(SS)}
\3tvec{p_b}{q_b}{r_b} \no \\
&+&(u_a,d_a,s_a)
\Mat3{m^2_{ab}(U^0U^0)}{m^2_{ab}(U^0D^0)}{0}
{m^2_{ab}(D^0U^0)}{m^2_{ab}(D^0D^0)}{0}
{0}{0}{m^2_{ab}(S^0S^0)}
\3tvec{u_b}{d_b}{s_b}.}
%%%
After imposing the constraints which are derived from the potential minimum
condition on the elements of matrices, we get
\eqn{
m^2_{ab}(U^+U^-) &=&m^2_{ab}(UU)=m^2_{ab}(U^0U^0) \no \\
&=&\Mat3{m^2_{H^U}+\lambda^2_{45}v^2_s}{0}{\epsilon^3M^2_{BU}c_c}
{0}{m^2_{H^U}+\lambda^2_{45}v^2_s}{\epsilon^3M^2_{BU}s_c}
{\epsilon^3M^2_{BU}c_c}{\epsilon^3M^2_{BU}s_c}{A_{23}v_sv_u/v_d},  \\
m^2_{ab}(D^+D^-)&=&m^2_{ab}(DD)=m^2_{ab}(D^0D^0) \no \\ 
&=&\Mat3{m^2_{H^D}+\lambda^2_{45}v^2_s-D_s}{0}{\epsilon^3M^2_{BD}c_c}
{0}{m^2_{H^D}+\lambda^2_{45}v^2_s-D_s}{\epsilon^3M^2_{BD}s_c}
{\epsilon^3M^2_{BD}c_c}{\epsilon^3M^2_{BD}s_c}{A_{23}v_sv_u/v_d} ,\\
m^2_{ab}(U^+D^-)&=&m^2_{ab}(UD)=-m^2_{ab}(U^0D^0) \no \\
&=&\Mat3{A_{45}v_s}{0}{\epsilon^3A_{67}v_sc_c}
{0}{A_{45}v_s}{\epsilon^3A_{67}v_ss_c}
{\epsilon^3A_{89}v_sc_c}{\epsilon^3A_{89}v_ss_c}{A_{23}v_s} ,\\
m^2(D^+U^-)&=&m^2_{ab}(DU)=-m^2_{ab}(D^0U^0)=[m^2_{ab}(U^+D^-)]^T \\
m^2_{ab}(SS)&=&
\Mat3{m^2_{S_1}+D_s}{0}{0}
{0}{m^2_{S_4}(s_s/c_s)}{-m^2_{S_4}}
{0}{-m^2_{S_4}}{m^2_{S_4}(c_s/s_s)} , \\
m^2_{ab}(S^0S^0)&=&
\Mat3{m^2_{S_1}+D_s}{0}{0}
{0}{18g^2_sc^2_sv^2_s+m^2_{S_4}(s_s/c_s)}{18g^2_sc_ss_sv^2_s-m^2_{S_4}}
{0}{18g^2_sc_ss_sv^2_s-m^2_{S_4}}{18g^2_ss^2_sv^2_s+m^2_{S_4}(c_s/s_s)} ,\\
&&D_s=9g^2_sv^2_s, \\
&&\lambda_{nm}=\lambda_nc_s+\lambda_ms_s, \\
&&A_{nm}=A_nc_s+A_ms_s , \\
&&M^2_{BU}=(\lambda_{23}\lambda_{67}+\lambda_{45}\lambda_{89})v^2_s-m^2_{BU} ,\\
&&M^2_{BD}=(\lambda_{23}\lambda_{89}+\lambda_{45}\lambda_{67})v^2_s-m^2_{BD} ,
}
where $O(v_{u,d})$ contributions are neglected. After the field redefinitions as
\eqn{
\3tvec{X_1}{X_2}{X_3}
&=&\Mat3{c_c}{-s_c}{0}
{s_c}{c_c}{0}
{0}{0}{1}
\3tvec{X'_1}{X'_2}{X'_3} ,\quad X=(H^U,H^D), \\
\3tvec{S_1}{S_2}{S_3}
&=&\Mat3{1}{0}{0}
{0}{c_s}{-s_s}
{0}{s_s}{c_s}
\3tvec{S'_1}{S'_2}{S'_3},
}
we get approximately diagonalized mass matrices
\eqn{
\L[m^2_{ab}(U^+U^-)\R]'
&=&\Mat3{m^2_{H^U}+\lambda^2_{45}v^2_s}{0}{\epsilon^3M^2_{BU}}
{0}{m^2_{H^U}+\lambda^2_{45}v^2_s}{0}
{\epsilon^3M^2_{BU}}{0}{A_{23}v_sv_d/v_u} ,\\
\L[m^2_{ab}(U^+D^-)\R]'
&=&\Mat3{A_{45}v_s}{0}{\epsilon^3A_{67}v_s}
{0}{A_{45}v_s}{0}
{\epsilon^3A_{89}v_s}{0}{A_{23}v_s} ,\\
\L[m^2_{ab}(D^+D^-) \R]'
&=&\Mat3{m^2_{H^D}+\lambda^2_{45}v^2_s-D_s}{0}{\epsilon^3M^2_{BD}}
{0}{m^2_{H^D}+\lambda^2_{45}v^2_s-D_s}{0}
{\epsilon^3M^2_{BD}}{0}{A_{23}v_sv_u/v_d}, \\
\L[m^2_{ab}(SS) \R]'
&=&\Mat3{m^2_{S_1}+D_s}{0}{0}
{0}{0}{0}
{0}{0}{m^2_{S_4}/c_ss_s} ,\\
\L[m^2_{ab}(S^0S^0) \R]'
&=&\Mat3{m^2_{S_1}+D_s}{0}{0}
{0}{18g^2_sv^2_s}{0}
{0}{0}{m^2_{S_4}/c_ss_s}.
}
In order to get the mass of the lightest neutral Higgs boson, we take account of $O(v_{u,d})$
contributions  and diagonalize the mass matrix by the field redefinition as
\eqn{
\L(
\begin{array}{c}
u_3 \\
d_3 \\
s_2 \\
s_3 \\
\end{array}
\R)
=
\L(
\begin{array}{cccc}
v_u/v &-v_d/v &0    &0 \\
v_d/v &v_u/v  &0    &0 \\
0       &0        &c_s  &-s_s \\
0       &0        &s_s  &c_s \\
\end{array}
\R)
\L(
\begin{array}{c}
\phi_1 \\
\phi_2 \\
\phi_3 \\
\phi_4 \\
\end{array}
\R),
}
then we get the squared masses of $\phi_{1,2,3,4}$ as follows: 
\eqn{
m^2_1&=&\frac12(g^2_Y+g^2_2)(v^2_u-v^2_d)^2/v^2
+4(\lambda^2_2+\lambda^2_3)v^2_uv^2_d/v^2=[0.0687+0.75(\lambda^2_2+\lambda^2_3)]v^2, \\
m^2_2&=&A_{23}v_sv^2/v_uv_d+O(v^2), \\
m^2_3&=&18g^2_sv^2_s, \\
m^2_4&=&m^2_{S_4}/c_ss_s,
}
where $g_Y=0.357, g_2=0.65$ are substituted into Eq.(141).
Here, $m_1$ is a tree level contribution to the mass of 
the lightest CP-even neutral Higgs boson $\phi_1$.
While the two contributions from $\lambda_2$ and $\lambda_3$ to $m_1$ are additive,
the contributions from them to the higgsino mass: $\lambda_{23}v_s$ could 
be destructive, and if so, which  enhances the lepton $g-2$ through reducing the higgsino mass.
The experimental value $125\mbox{GeV}$ is realized by adding 
the one-loop contribution from the stop \cite{SUSY-Higgs}
\eqn{
\Delta m^2_1\simeq \frac{3m^2_t}{4\pi^2}\ln\frac{m^2_T}{m^2_t},
}
where $m_T$ is stop mass. The one-loop contributions from the G Higgses 
assist to push up the Higgs mass \cite{u1-Higgs}.

\subsection{G Higgs}

The mass terms for the G Higgses are derived from the superpotential:
\eqn{
W=k_2S_2G_aG^c_a+k_3S_3G_aG^c_a,
}
from which mass terms are given by
\eqn{
V&\supset&m^2_G|G_a|^2+m^2_{G^c}|G^c|^2 -A_{k23}v_sG_aG^c_a+h.c. \no \\
&+&|k_2G_aG^c_a+\lambda_2v_uv_d|^2
+|k_3G_aG^c_a+\lambda_3v_uv_d|^2+k^2_{23}(|v_sG|^2+|v_sG^c|^2)+\mbox{D-terms} , \\
&&A_{k23}=A_{k2}c_s+A_{k3}s_s, \quad k_{23}=k_2c_s+k_3s_s.
}
Due to the $S_4$ symmetry, the triplets $G,G^c$ have unified mass matrices 
as follows:
\eqn{
V&\supset&(G^*_a,G^c_a)M^2_G
\2tvec{G_a}{(G^c_a)^*} ,\\
M^2_G&=&
\mat2{m^2_G+(k_{23}v_s)^2-\frac13m^2_{Z'}}
{(k_2\lambda_2+k_3\lambda_3)v_uv_d-A_{k23}v_s}
{(k_2\lambda_2+k_3\lambda_3)v_uv_d-A_{k23}v_s}{m^2_{G^c}+(k_{23}v_s)^2-\frac16m^2_{Z'}}.
}
The interaction terms for the G-Higgses are given by
\eqn{
W_G=GQY^{QQ}Q+G^cU^cY^{UD}D^c+GU^cY^{UE}E^c+G^cQY^{QL}L+GD^cY^{DN}N^c,
}
where the coupling matrices are given by
\eqn{
&&Y^{QQ}=\Mat3{\epsilon^{21}}{\epsilon^{21}}{\epsilon^{19}}
{\epsilon^{21}}{\epsilon^{21}}{\epsilon^{19}}
{\epsilon^{19}}{\epsilon^{19}}{\epsilon^{17}} , \quad
Y^{UD}=\Mat3{\epsilon^{28}}{\epsilon^{27}}{\epsilon^{27}}
{\epsilon^{25}}{\epsilon^{24}}{\epsilon^{24}}
{\epsilon^{24}}{\epsilon^{23}}{\epsilon^2}, \\
&&Y^{UE}=\Mat3{\epsilon^{24}}{\epsilon^{22}}{\epsilon^{42}}
{\epsilon^{21}}{\epsilon^{36}}{\epsilon^{39}}
{\epsilon^{20}}{\epsilon^{35}}{\epsilon^{38}} , \quad
Y^{QL}=\Mat3{\epsilon^{25}}{\epsilon^{25}}{\epsilon^{25}}
{\epsilon^{25}}{\epsilon^{25}}{\epsilon^{25}}
{\epsilon^{23}}{\epsilon^{23}}{\epsilon^{23}}, \quad
Y^{DN}=\Mat3{\epsilon^{24}}{\epsilon^{24}}{\epsilon^{24}}
{\epsilon^{23}}{\epsilon^{23}}{\epsilon^{23}}
{\epsilon^2}{\epsilon^2}{\epsilon^2}.
}
As $Y^{QQ},Y^{UE},Y^{QL}$ do not cause any observable effect, 
they are out of our consideration.
At the SCKM basis, $Y^{UD},Y^{DN}$ are redefined as follows:
\eqn{
(Y^{UD})_{\mbox{SCKM}}&=&R^T_UV^T_K(U)Y^{UD}V_K(D)R_D
=\Mat3{\epsilon^{17}}{\epsilon^{16}}{\epsilon^{12}}
{\epsilon^{14}}{\epsilon^{13}}{\epsilon^{9}}
{\epsilon^{7}}{\epsilon^{6}}{\epsilon^{2}} , \\
(Y^{DN})_{\mbox{SCKM}}&=&R^T_DV^T_K(D)(Y^{DN})V_K(N)
=\Mat3{\epsilon^7}{\epsilon^7}{\epsilon^7}
{\epsilon^6}{\epsilon^6}{\epsilon^6}
{\epsilon^2}{\epsilon^2}{\epsilon^2}.
}
As the large elements in the $(3,3)$-entry of $(Y^{UD})_{\mbox{SCKM}}$ 
and the third row of $(Y^{DN})_{\mbox{SCKM}}$ induces very fast decay of $G,G^c$,
the success of BBN is not spoiled.
Further more, if $G\to n+b$ is the dominant decay mode of $G$,
we can verify the RHN directly  at a collider experiment.
These couplings also open the dangerous  channel to the proton decay.
The dominant contributions to the proton decay are induced by the couplings
\eqn{
(Y^{us})(Y^{dn})=(\epsilon^{16})(\epsilon^7)=\epsilon^{23}, \quad
(Y^{ud})(Y^{sn})=(\epsilon^{17})(\epsilon^6)=\epsilon^{23}.
}
Multiplied by the factor
\eqn{
\frac{\epsilon^6v_u}{M_R}\sim \epsilon^6,
}
which comes from $N^c-\nu$ mixing, the dimension-less coefficient of the 4-Fermi operator
\eqn{
{\cal L} \supset \frac{c_{uds\nu}}{M^2(G)}\bar{u}d\bar{\nu}s ,
}
is estimated  as 
\eqn{
c_{uds\nu}\sim \epsilon^{29},
}
which is consistent with the experimental bound for $p\to K^++\nu$:
\eqn{
c_{uds\nu}< 10^{-27}.
}

As the single G interactions violate $B+L$ while they conserve $B-L$,
they may assist the $B+L$ violating process which converts a lepton number to
a baryon number. 
The dominant terms which contribute to this process  are given by
\eqn{
W&=&\epsilon^2Y^{UD}_3(G^c_1+G^c_2+G^c_3)U^c_3D^c_3 \no \\
&+&\epsilon^2Y^{DN}_1[\sqrt{3}(G_2-G_3)N^c_1+(G_1+G_2-2G_3)N^c_2]D^c_3 \no \\
&+&\epsilon^8Y^{DN}_1(G_1+G_2+G_3)D^c_3(c_NN^c_1+s_NN^c_2).
}
The contributions from the first line and the second line in Eq.(161) to the term
\eqn{
(\epsilon^2Y^{UD}_3)(\epsilon^2Y^{DN}_1)U^c_3D^c_3N^c_iD^c_3,
}
is canceled due to the mass degeneracy in G Higgs \cite{proton}.
As any linear combinations of $G_a$ which are gotten by
unitary transformation are assumed to be mass eigenstates,
we can move to more convenient view point.
Up to $O(\epsilon^2)$, since we can assign the $S_3$ singlet
\eqn{
G_D=\frac{G_1+G_2+G_3}{\sqrt{3}}
}
to a diquark and the $S_3$ doublet
\eqn{
G_{Li}=\L(\frac{G_2-G_3}{\sqrt{2}},\frac{G_1+G_2-2G_3}{\sqrt{6}}\R)
}
to a leptoquark, the baryon number and the lepton number are conserved respectively.
However they are violated by including the $O(\epsilon^8)$ terms.
The contribution to Eq.(162) is induced by the first line and third line
in Eq.(161).
Requiring the process $n+\bar{t}\to b+b$ is in equilibrium, we get the constraint
\eqn{
1<\frac{\Gamma(n+\bar{t}\to b+b)}{H(m_N)}
\sim 10^{12}\epsilon^{20}(Y^{UD}_3Y^{DN}_1)^2\frac{m^4_N}{m^4_G}\sim 10^{-8}(Y^{UD}_3Y^{DN}_1)^2,
}
which is difficult to be satisfied. Therefore the terms in Eq.(161) do not have
significant impact on leptogenesis.

\subsection{LSP}

As the R-parity is conserved in this model, the LSP is stable. 
We identify the singlino $s_1$ as the LSP which has a tiny mass
\eqn{
m(s_1)\sim \frac{(\epsilon^6v_u)(\epsilon^6v_d)}{\lambda_{23}v_s}
\sim 10^{-2}\mbox{eV}.
}
Although $s_1$ is not the dominant component of dark matter,
it may help to explain the delay of structure formation \cite{SF}.
Further more, $s_1$ behaves as an extra neutrino,  which changes  the effective number of
neutrinos to  (\cite{singlino})
\eqn{N_{eff}=3.097,}
where $m_{Z'}<4700\mbox{GeV}$ is assumed.
This extra contribution softens the Hubble tension 
between the distance ladder method \cite {H-ladder} and the CMB data \cite{H-CMB}.

The interaction of the bino which is the LSP of MSSM is given by
\eqn{
{\cal L}\supset ig_Y\frac{\epsilon^6v_d}{m_{\mbox{SUSY}}}(H^U_3)^*\lambda_Ys_1,
}
from which the bino lifetime is calculated as follows:
\eqn{
\Gamma(\lambda_T\to H+s_1)\sim
 \alpha_Ym_{\mbox{SUSY}}\L(\frac{\epsilon^6v_d}{m_{\mbox{SUSY}}}\R)^2
 \sim 10^{-4}\mbox{eV}&\to& 
\tau(\lambda_Y)\sim 10^{-11}\mbox{sec},
}
which is consistent with the standard cosmology.
The NLSP in our model is the lighter of two linear combinations of  two singlinos $s_{2,3}$ 
which must be heavier than 
$100\mbox{MeV}$ to avoid the longer lifetime than $1\mbox{sec}$ \cite{singlino-2}.
It is easy to give such a tiny mass to the NLSP.

%%%%%%%%%%%%%%%%%%%%%%%%%%%%%%%%%%%%%%%%%%%%%%%
% sec6
%%%%%%%%%%%%%%%%%%%%%%%%%%%%%%%%%%%%%%%%%%%%%%%

\section{Lepton anomalous magnetic dipole moments}

Here, we evaluate the lepton anomalous magnetic moments.
The $(g-2)_\mu$ has $3.7\sigma$ discrepancy between the SM prediction
and the experimental value as given in Eq.(1).
For the $(g-2)_e$, there is $2.4\sigma$
discrepancy between the new SM prediction and the experimental value
as given in Eq.(5).
These gaps are filled by the SUSY contributions. 
While the flavor blind contribution gives Eq.(6),
the experimental observation Eq.(7) does not obey it.
This discrepancy reflects non-trivial flavor structure of new physics.
In our model, this comes from the structure of charged lepton Yukawa matrices.

In the basis that the charged lepton and Higgs mass matrices are diagonalized,
the Yukawa interactions of the charged lepton  are given by
\eqn{
W_{E}&=&H^D_AL
\Mat3{\epsilon^2Y^{EI}_1}{\epsilon^{6}}{\epsilon^{7}}
{\epsilon^{8}}{Y^{EI}_2}{\epsilon^{7}}
{\epsilon^{8}}{\epsilon^{6}}{\epsilon^{5}}E^c
+H^D_BL
\Mat3{\epsilon^8}{-Y^{EI}_2}{\epsilon^7}
{\epsilon^2Y^{EI}_1}{\epsilon^6}{\epsilon^7}
{\epsilon^8}{\epsilon^6}{\epsilon^{13}}E^c \no \\
&+&H^D_3L
\Mat3{\epsilon^5Y^E_1}{\epsilon^{15}}{\epsilon^{8}}
{\epsilon^{17}}{\epsilon^3Y^E_2}{\epsilon^{8}}
{\epsilon^{11}}{\epsilon^{9}}{\epsilon^2Y^E_3}
E^c ,
}
where $H^D_A=(H^D_1)'$ and $H^D_B=(H^D_2)'$ are mass eigenstates defined by Eq.(133).
These interactions induce $\mu\to e+\gamma$ process and the
experimental constraint for the branching ratio
\eqn{
BR(\mu\to e\gamma)
=\frac{48\pi^3\alpha_{em}}{G^2_F}\L(\frac{\epsilon^6Y^{IE}_2}{192\pi^2m^2_{A,B}}\R)^2
<4.2\times 10^{-13},
}
which gives the lower mass bound for $H^D_{A,B}$ as
\eqn{
\frac{m_{A,B}}{\sqrt{Y^{EI}_2}}>15\mbox{GeV},
}
where $\alpha_{em}=1/137$ and $G_F=1.166\times 10^{-5}\mbox{GeV}^{-2}$
are used. This constraint is easily satisfied.

The chargino and the neutralino mass matrices  are given by
\eqn{
{\cal L}&=&-\chi^T_+ M_C \chi_- 
-\frac12 \chi^TM_N\chi
-\lambda_{45}v_s(h^U_B)^+(h^D_B)^-
-\lambda_{45}v_s(h^U_B)^0(h^D_B)^0
+h.c. ,}
%%%
\eqn{M_C&=&\Mat3{\lambda_{45} v_s}{\epsilon^3\lambda_{67}v_s}{g_2N_U\epsilon^3v_u}
{\epsilon^3\lambda_{89}v_s}{\lambda_{23} v_s}{g_2v_u}
{g_2N_D\epsilon^3v_d}{g_2v_d}{M_2} , }
%%%
\eqn{M_N&=&
\L(
\begin{array}{cccccc}
0  & 0 & \lambda_{45}v_s & \epsilon^3\lambda_{67}v_s &\epsilon^3N_Ug_Yv_u/\sqrt{2} 
&-\epsilon^3N_Ug_2v_u/\sqrt{2} \\
** & 0 & \epsilon^3\lambda_{89}v_s & \lambda_{23}v_s              &g_Yv_u/\sqrt{2}                   
&-g_2v_u/\sqrt{2} \\
** & * & 0                    &0                                  &-\epsilon^3N_Dg_Yv_d/\sqrt{2} 
&\epsilon^3N_Dg_2v_d/\sqrt{2} \\
** & * & *                    & 0                                 &-g_Yv_d/\sqrt{2}                    
& g_2v_d/\sqrt{2} \\
** & * & *                    & *                                 &-M_Y                                        
&0 \\
** & * & *                    & *                                 & *                                       
&-M_2\\
\end{array}
\R)=M^T_N ,\\
\chi^T_-&=&((h^D_A)^-,(h^D_3)^-,w^-), \\
\chi^T_+&=&((h^U_A)^+,(h^U_3)^+,w^+), \\
w^\pm&=&\frac{\mp i\lambda^1_2 -\lambda^2_2}{\sqrt{2}}, \\
\chi^T&=&(h^U_A,h^U_3,h^D_A,h^D_3,i\lambda_Y,i\lambda^3_2) ,
}
which are diagonalized by bi-unitary translation and unitary translation respectively as
\eqn{
&&\chi_+=U\chi'_+, \quad \chi_-=D\chi'_-, \quad 
U^T M_C D =\mbox{diag}(\mu_1,\mu_2,\mu_3), \\
&&\chi=V\chi', \quad
V^TM_NV=\mbox{diag}(\xi_1,\xi_2,\cdots, \xi_6),
}
from which we calculate the one-loop contributions to the muon and electron  $g-2$ as follows:
\eqn{
a_{\mu,e}(\mbox{SUSY})&=&a_{\mu,e}(\chi^\pm)+a_{\mu,e}(\chi^0)+a_{\mu,e}(h_B) , \\
a_\mu(\chi^\pm)&=&\frac{m_\mu}{16\pi^2m^2(N)}\sum_{a=1,2,3}
\L\{\frac13 m_\mu (|C^L_{\mu,a}|^2+|C^R_{\mu,a}|^2)f_C(x_a)
-3\mu_a\mbox{Re}[C^L_{\mu,a}(C^R_{\mu,a})^*]g_C(x_a) \R\} ,\no\\ \\
a_e(\chi^\pm)&=&\frac{m_e}{16\pi^2m^2(N)}\sum_{a=1,2,3}
\L\{\frac13 m_e (|C^L_{e,a}|^2+|C^R_{e,a}|^2)f_C(x_a)
-3\mu_a\mbox{Re}[C^L_{e,a}(C^R_{e,a})^*]g_C(x_a) \R\} ,\no \\}
%%%
\eqn{f_C(x)&=&\frac{1}{(1-x)^4}\L(1+\frac32 x-3x^2+\frac12 x^3+3x\ln x\R), \\
g_C(x)&=&\frac{1}{(1-x)^3}\L(1-\frac43 x+\frac13 x^2+\frac23\ln x\R), \\
x_a&=&\frac{\mu^2_a}{m^2(N)},  \quad m^2(N)=m^2_L+\frac{g^2_Y+g^2_2}{4}(v^2_d-v^2_u), }
%%%
\eqn{C^L_{\mu,a}&=&Y^{EI}_2D_{1a}+\epsilon^3Y^E_2D_{2a} ,\\
C^R_{\mu,a}&=&-g_2U_{3a} , }
%%%
\eqn{C^L_{e,a}&=&\epsilon^2Y^{EI}_1D_{1a}+\epsilon^5Y^E_1D_{2a} , \\
C^R_{e,a}&=&-g_2U_{3a}, }

%%%
\eqn{a_\mu(\chi^0)&=&-\frac{m_\mu}{16\pi^2m^2(E)}
\sum^6_{a=1}\L\{
\frac16 m_\mu(|N^L_{\mu,a}|^2+|N^R_{\mu,a}|^2)f_N(y_a)
+\xi_a\mbox{Re}[N^L_{\mu, a}(N^R_{\mu, a})^*]g_N(y_a) \R\} ,\no\\ \\
a_e(\chi^0)&=&-\frac{m_e}{16\pi^2m^2(E)}
\sum^6_{a=1}\L\{
\frac16 m_e(|N^L_{e,a}|^2+|N^R_{e,a}|^2)f_N(y_a)
+\xi_a\mbox{Re}[N^L_{e,a}(N^R_{e,a})^*]g_N(y_a)\R\}, \no\\ \\
f_N(x)&=&\frac{1}{(1-x)^4}(1-6x+3x^2+2x^3-6x^2\ln x), \\
g_N(x)&=&\frac{1}{(1-x)^3}(1-x^2+2x\ln x) ,}
%%%
\eqn{y_a&=&\frac{\xi^2_a}{m^2(E)}, \quad m^2(E)=m^2_L+\frac{g^2_Y-g^2_2}{4}(v^2_d-v^2_u), }
%%%
\eqn{N^L_{\mu,a}&=&-Y^{EI}_2V_{3a}-\epsilon^3Y^E_2V_{4a} ,\\
N^R_{\mu,a}&=&\frac{g_2}{\sqrt{2}}V_{6a}+\frac{g_Y}{\sqrt{2}}V_{5a}, \\
N^L_{e,a}&=&-\epsilon^2Y^{EI}_1V_{3a}-\epsilon^5Y^E_1V_{4a} , \\
N^R_{e,a}&=&\frac{g_2}{\sqrt{2}}V_{6a}+\frac{g_Y}{\sqrt{2}}V_{5a}, }
%%%
\eqn{a_\mu(h_B)&=&\frac{m^2_\mu |Y^{EI}_2|^2}{16\pi^2}
\L(\frac{f_C(x_B)}{3m^2(N)}-\frac{f_N(y_B)}{6m^2(E)}\R), \\
a_e(h_B)&=&\frac{m^2_e |\epsilon^2Y^{EI}_1|^2}{16\pi^2}
\L(\frac{f_C(x_B)}{3m^2(N)}-\frac{f_N(y_B)}{6m^2(E)}\R), \\
x_B&=&\frac{|\lambda_{45}v_s|^2}{m^2(N)}, \quad y_B=\frac{|\lambda_{45}v_s|^2}{m^2(E)},
}
where $m_L$ is the $S_4$-doublet left-handed slepton mass. 
In calculating the neutralino contributions, 
we omitted the negligible  contributions from
the right-handed slepton: $E^c_{1,2}$, the singlino, and the $\mbox{U}(1)_S$ gaugino.

At the degenerated mass and large $N_U$ limit:
\eqn{
m_L=\mu_1=\mu_2=\mu_3, \quad N_U\gg 1,
}
then we get
\eqn{
a_\mu&=&\frac{g^2_2\epsilon^3Y^{EI}_2N_Uv_u m_\mu}{32\pi^2m^2_L}
=27\times 10^{-10}\L(\frac{200\mbox{GeV}}{m_L}\R)^2
\L(\frac{N_U}{13}\R)\L(\frac{Y^{EI}_2}{0.4}\R), \\
\frac{m^2_\mu a_e}{m^2_ea_\mu }
&=&\frac{\epsilon^2Y^{EI}_1m_\mu}{Y^{EI}_2 m_e}
\simeq \frac{2Y^{EI}_1}{Y^{EI}_2}.
}
The experimental values given in Eq.(1) and Eq.(5)
are realized by putting  by hand as
\eqn{
m_L=200 \mbox{GeV} ,\quad N_U=13, \quad Y^{EI}_1=-2.8, \quad Y^{EI}_2=0.4.
}
Assuming $N_D=1$ and imposing Eq.(96), we get
\eqn{
N_D=1, \quad Y^E_1=3.36, \quad Y^E_2=0.79.
}
The values of coupling constants at the Planck scale:
\eqn{
Y^E_1(M_P)=1.77, \quad Y^{EI}_1(M_P)=-1.47, \quad 
Y^E_2(M_P)=0.42, \quad Y^{EI}_2(M_P)=0.21,
}
are consistent with the $O(1)$ criterion.
The enhancement of the $(g-2)_e$ compared to the $(g-2)_\mu$
is originated from a large cancellation between two terms
in the electron mass: $m_e=(Y^E_1+Y^{EI}_1N_D)\epsilon^5 v_d$.

We give the numerical estimations of both $g-2$  as follows.
We define three parameter sets:
\eqn{
\mbox{Model A}&:& 200< \mbox{min}(|M_2|,|\lambda_{23}v_s|)<1000\mbox{GeV} , \quad 
\mbox{max}(|M_2|,|\lambda_{23}v_s|)=1.1\times \mbox{min}(|M_2|,|\lambda_{23}v_s|), \no \\
&& |\lambda_{45}v_s|=2\times \mbox{min}(|M_2|,|\lambda_{23}v_s|), \\
\mbox{Model B}&:&200< |\lambda_{45}v_s|<1000\mbox{GeV} , \quad
\mbox{min}(|M_2|,|\lambda_{23}v_s|)=1.2|\lambda_{45}v_s|, \no \\
&& \mbox{max}(|M_2|,|\lambda_{23}v_s|)=1.4|\lambda_{45}v_s|, \\
\mbox{Model C}&:&700< |M_2|, |\lambda_{23}v_s|, |\lambda_{45}v_s| <1000\mbox{GeV},
}
and the common parameter set:
\eqn{
&&200< |\lambda_{67}v_s|, |\lambda_{89}v_s|,  m_L<1000\mbox{GeV}, \quad  |M_Y|= 0.5|M_2|, \\
&&0.2<|Y^E_{1,2}|<3.0, \quad 0.2<|Y^{EI}_1|<3.0, \quad 0.2<|Y^{EI}_2|<0.4, \quad 0.5<|N_{U,D}|<10.
}
Taking account of the RGE factor
\eqn{
\frac{Y^E(M_S)}{Y^E(M_P)}=1.9,
}
for $Y^E_{1,2},Y^{EI}_1$, we have imposed the $O(1)$ criterion on the Yukawa couplings at $M_P$. 
In Model A, as the extra higgsinos are decoupled,  the advantage of large $Y^{EI}_{1,2}$
is not available unlike in the cases of Model B and Model C.
In Model C, an accidental degeneracy of the mass parameters could enhance the mixing angle of 
the higgsinos,
which is prevented in Model A and Model B. Note that such a enhancement needs
fine-tuning and so is unnatural.
We focus on the lightest charged SUSY particle in the loop
whose mass is defined as
\eqn{
\mu_L=\mbox{min}(|\mu_1|,|\mu_2|,|\mu_3|,m(E)).
}
\if0
\begin{center}
\begin{figure}
\includegraphics[width=5cm,pagebox=cropbox,clip]{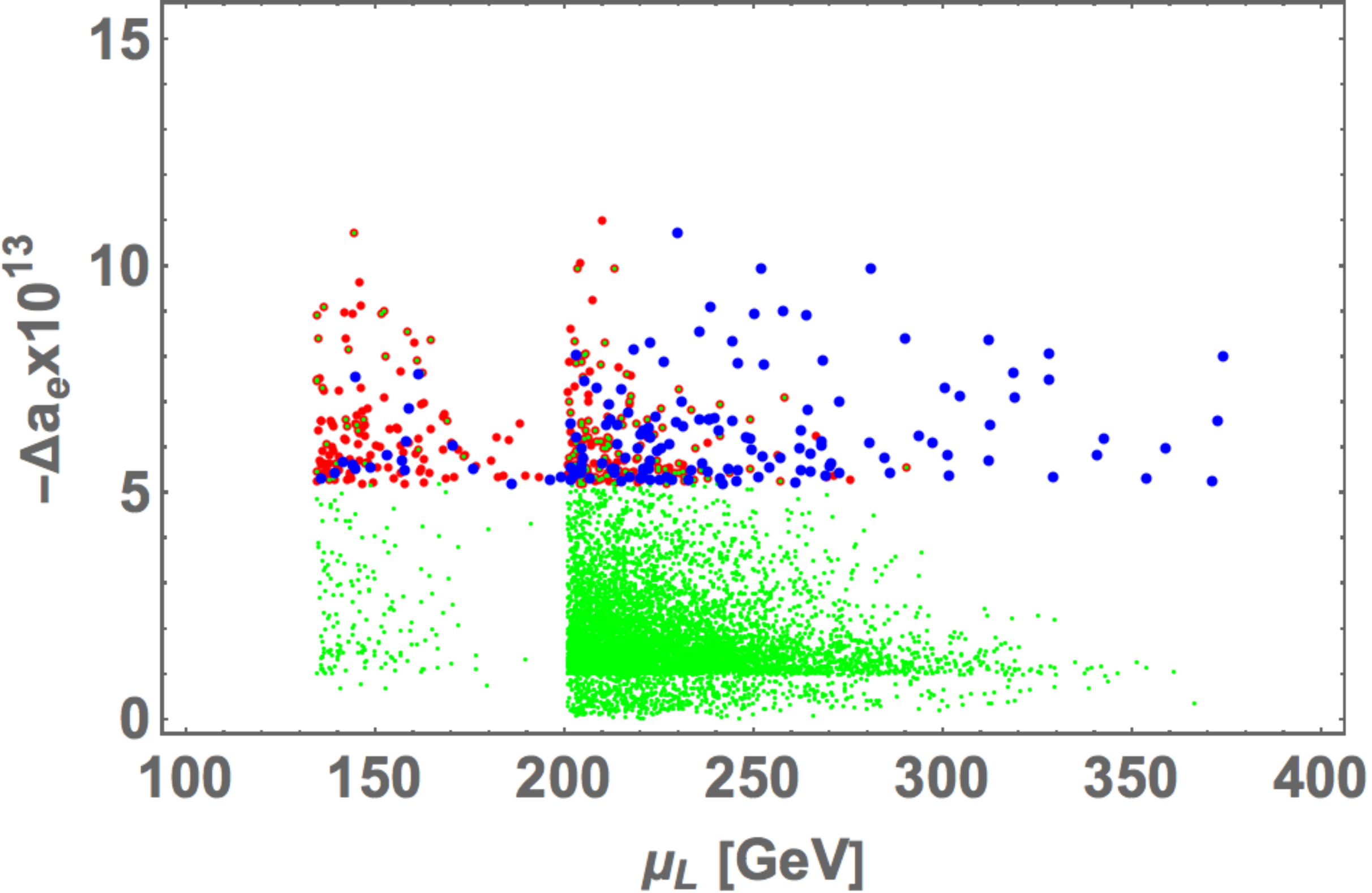}
\includegraphics[width=5cm,pagebox=cropbox,clip]{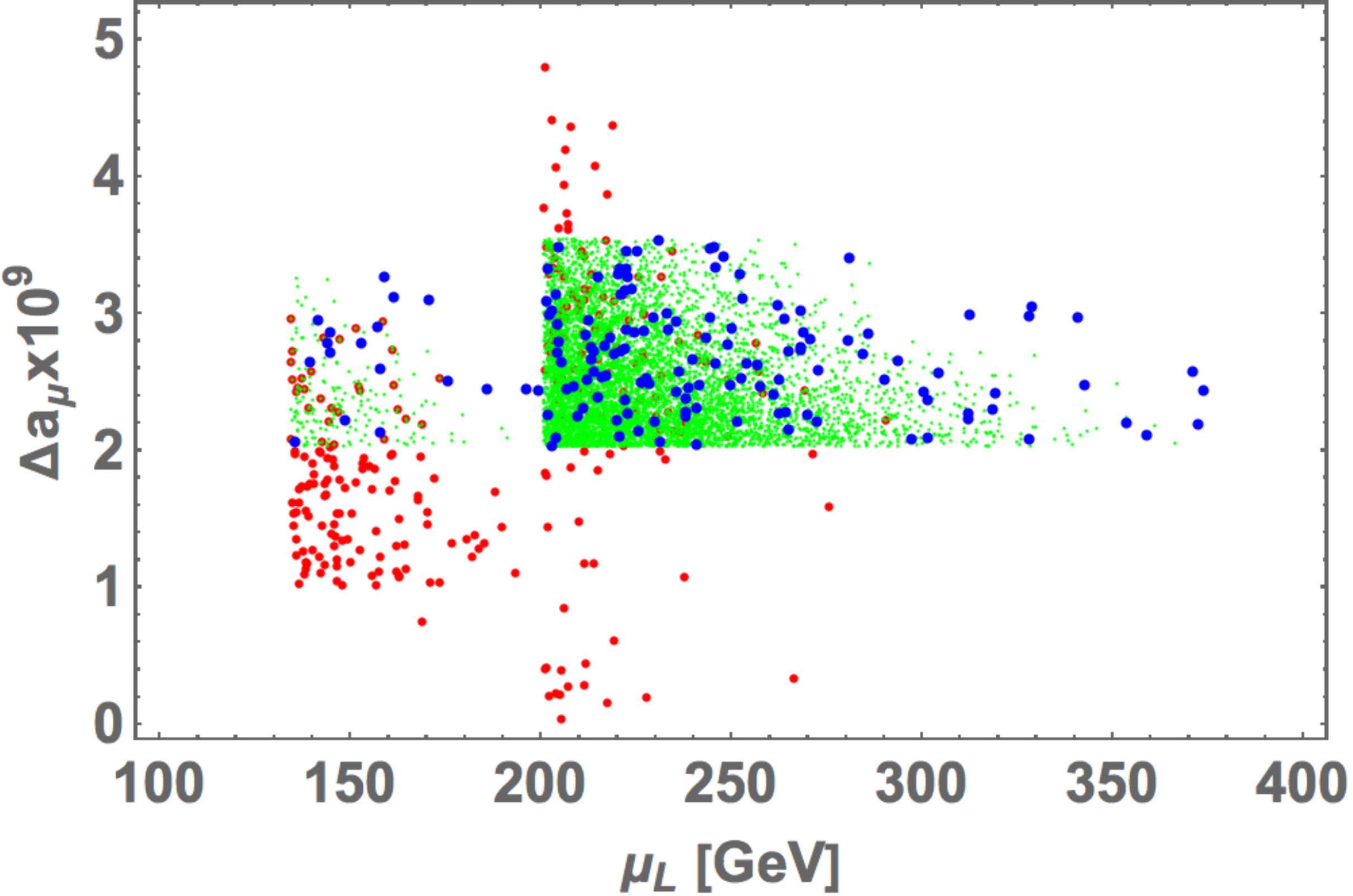}
\includegraphics[width=5cm,pagebox=cropbox,clip]{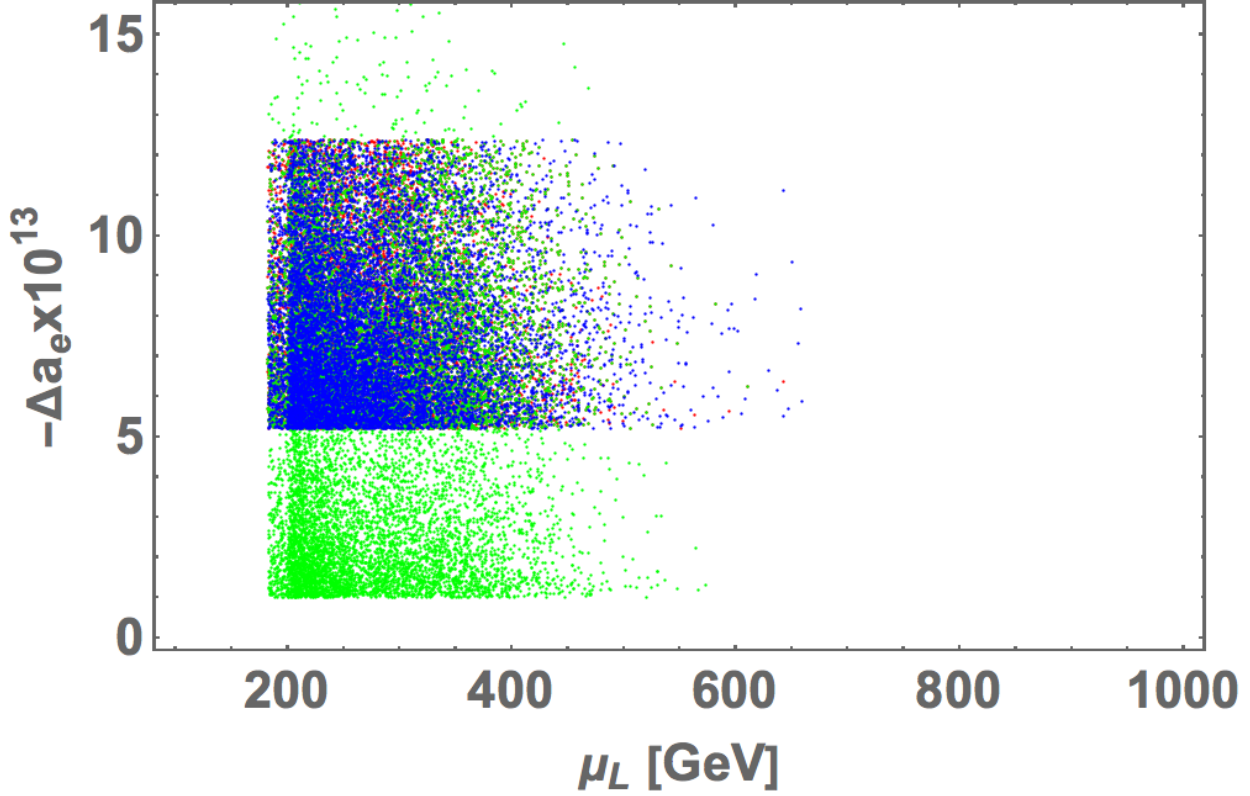}
\includegraphics[width=5cm,pagebox=cropbox,clip]{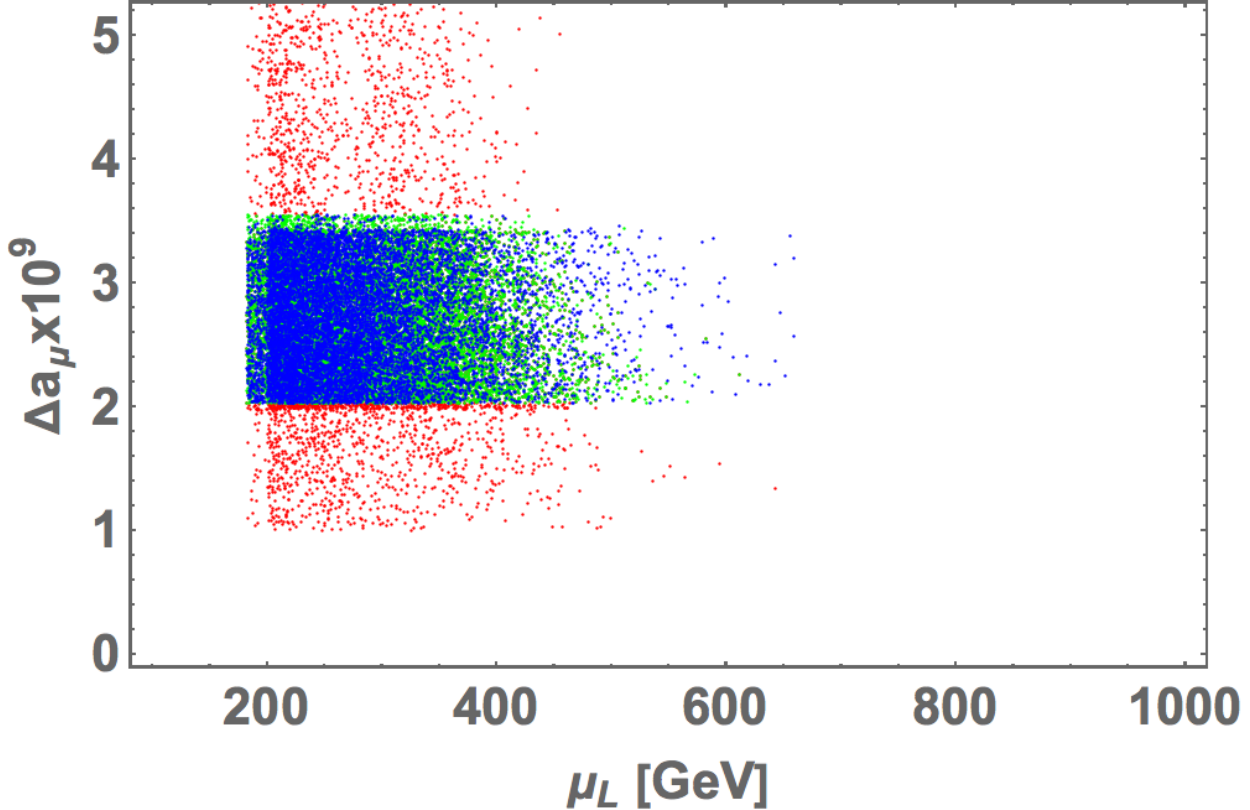}
\includegraphics[width=5cm,pagebox=cropbox,clip]{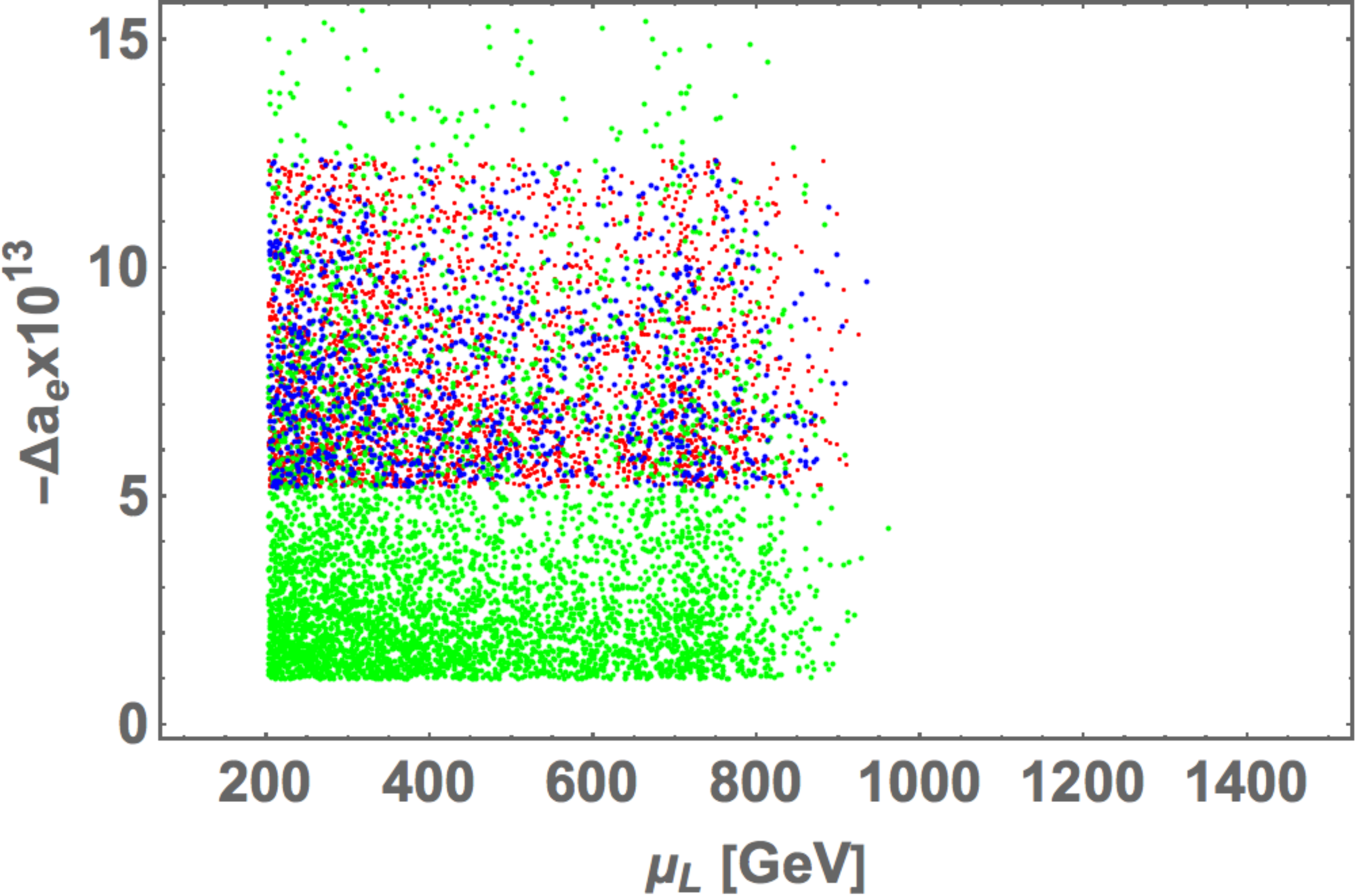}
\hspace{3cm}
\includegraphics[width=5cm,pagebox=cropbox,clip]{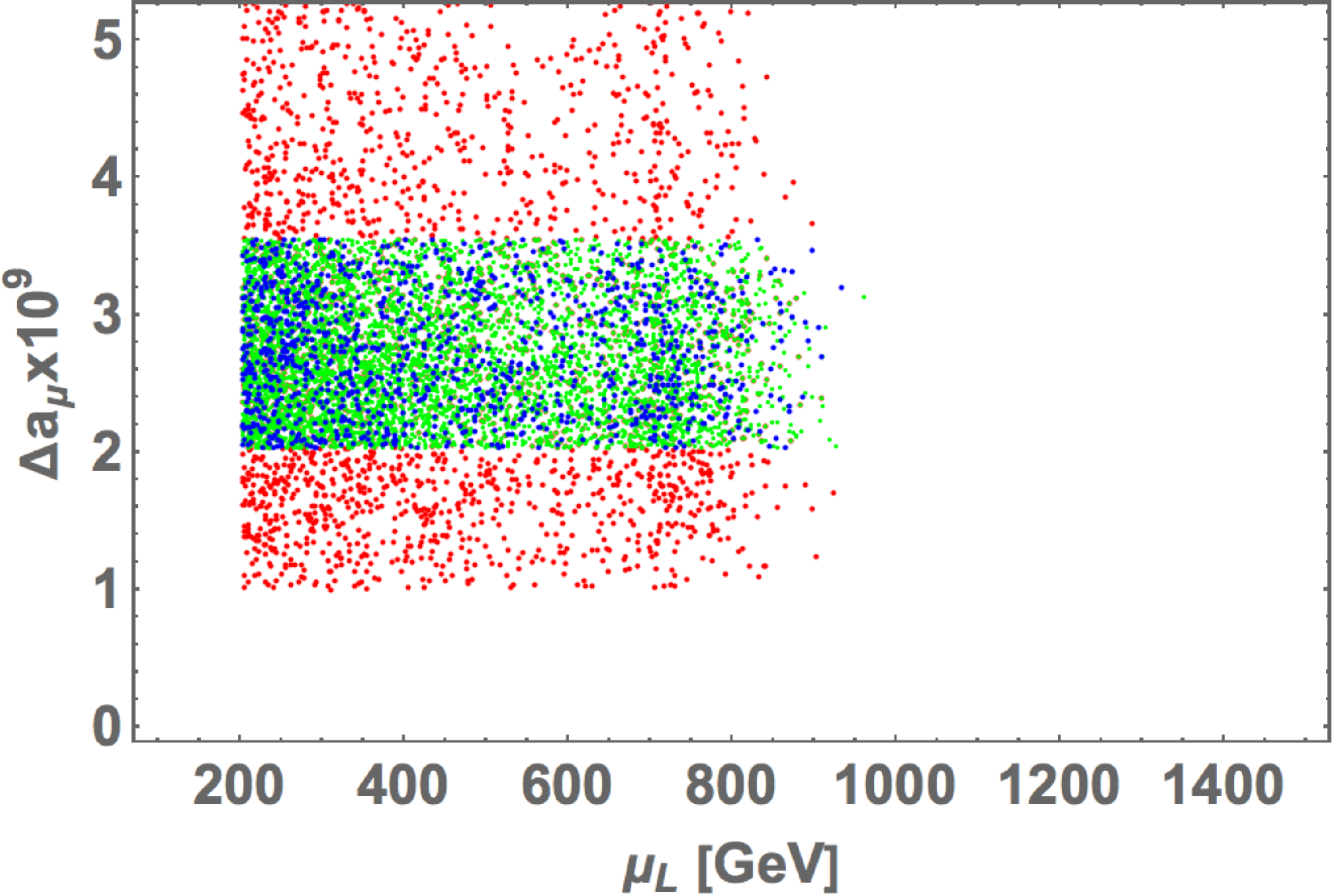}
\caption{Lepton anomalous magnetic moments of electron (left) and muon (right) 
for Model A(top), Model  B(middle) and Model C(bottom) .
Blue points satisfy both Eq.(1) and Eq.(5),  green points satisfy only Eq.(1),
and red points satisfy only Eq.(5).}
\end{figure}
\end{center}
\fi
%%%%%%%%%%%%%%%%%%%
\begin{figure}[tb!]\begin{center}
\includegraphics[width=80mm]{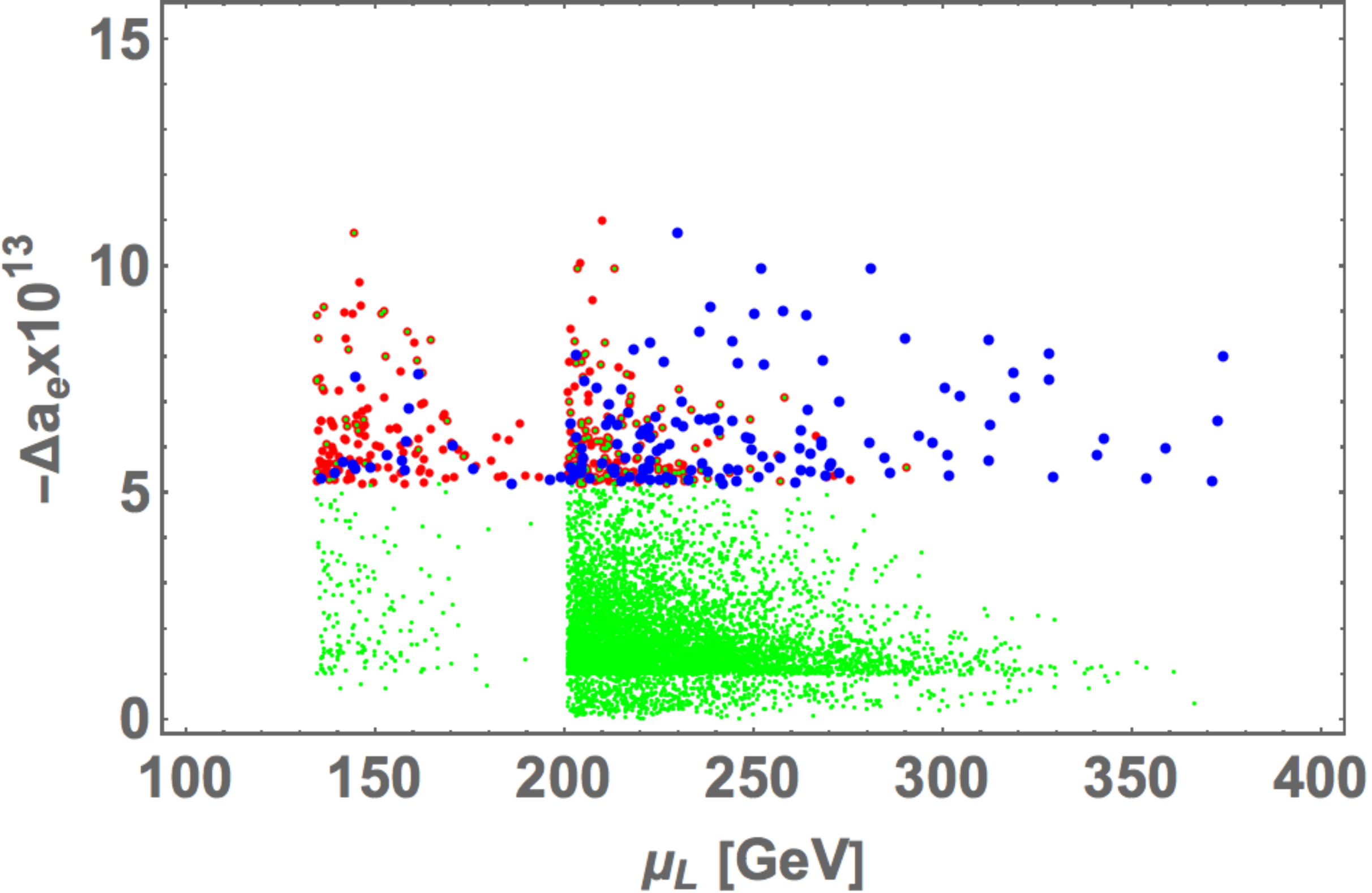} \ 
 \includegraphics[width=80mm]{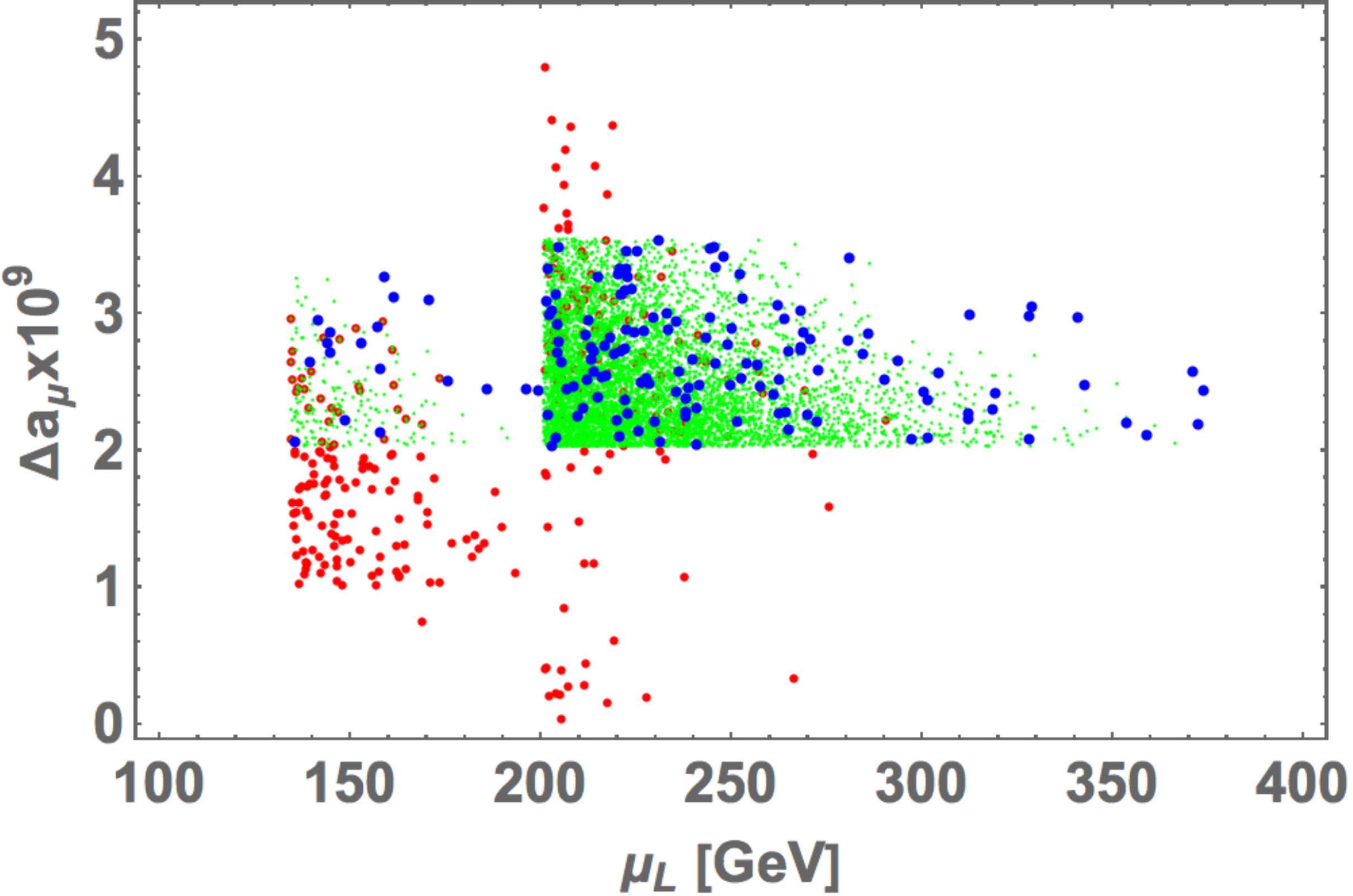}
  \includegraphics[width=80mm]{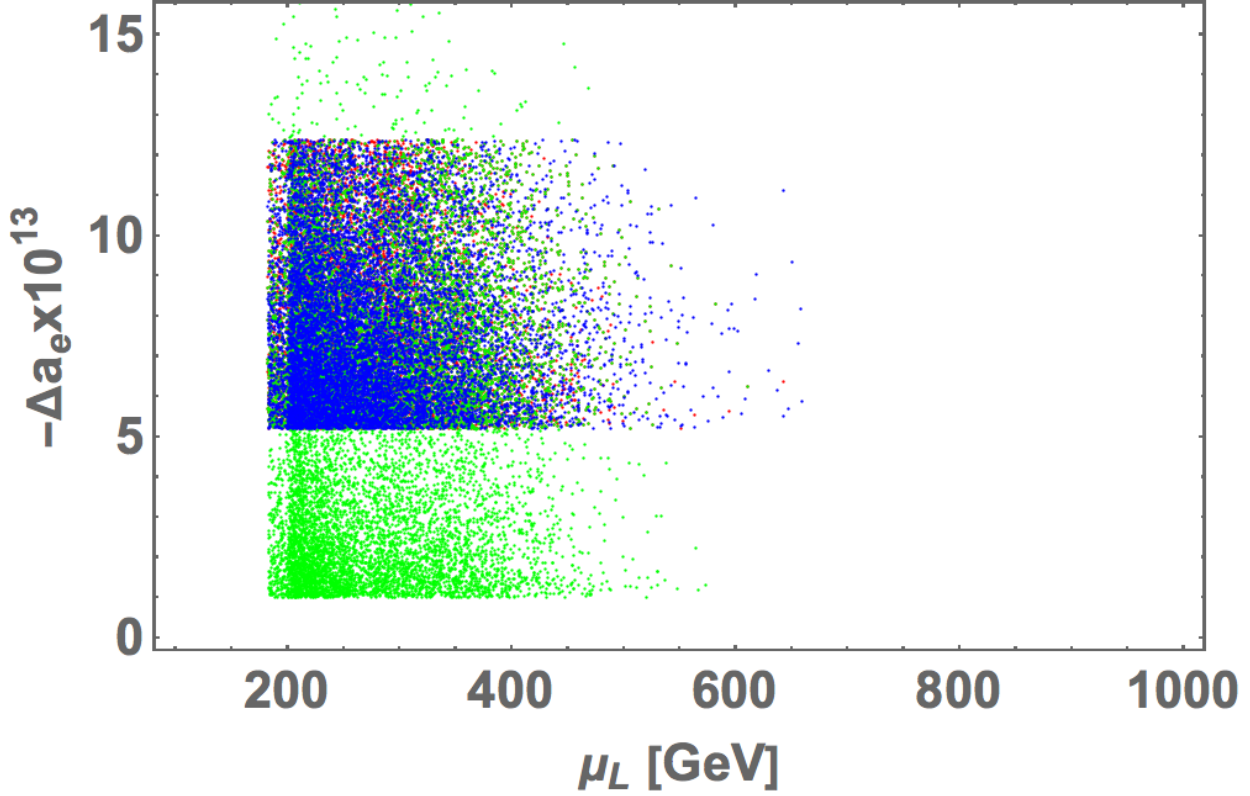} \ 
 \includegraphics[width=80mm]{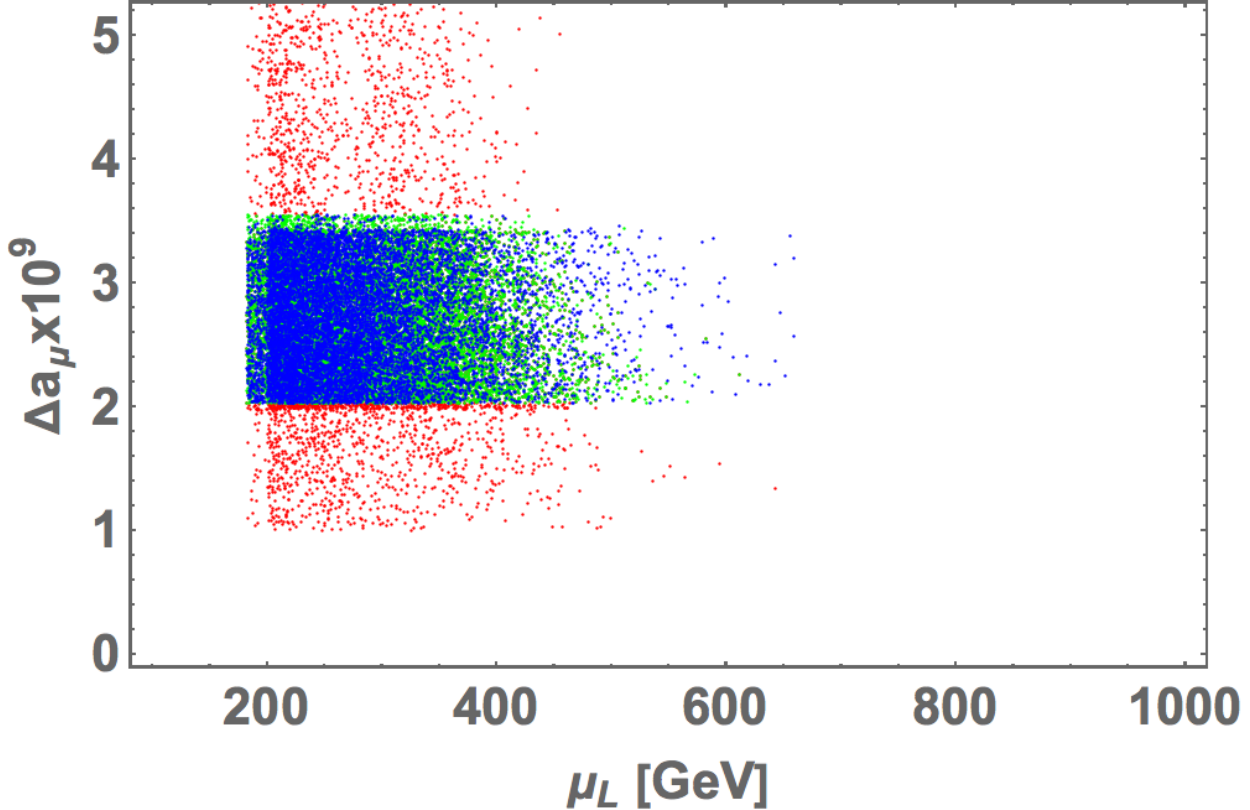}
\includegraphics[width=80mm]{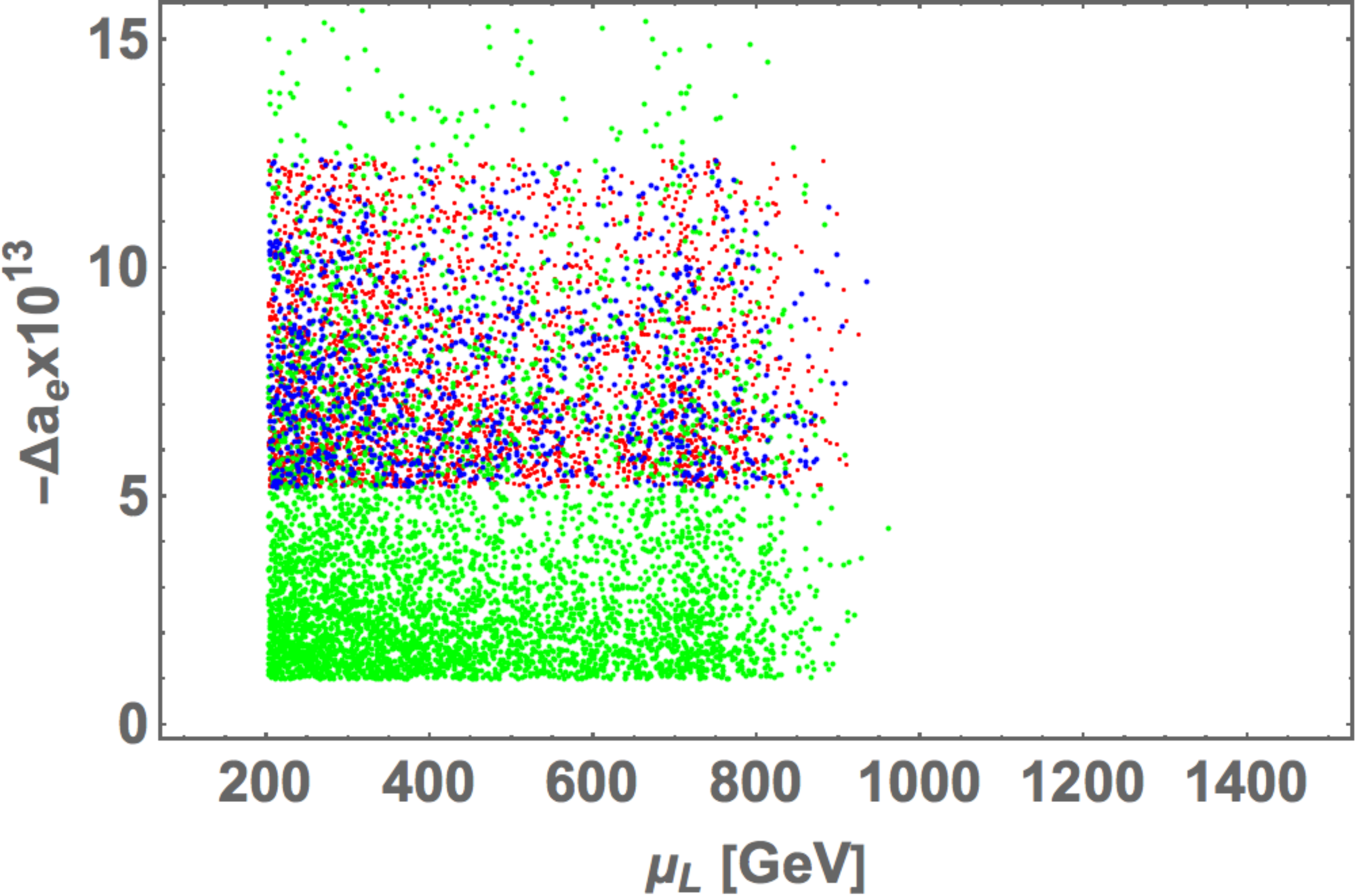} \ 
 \includegraphics[width=80mm]{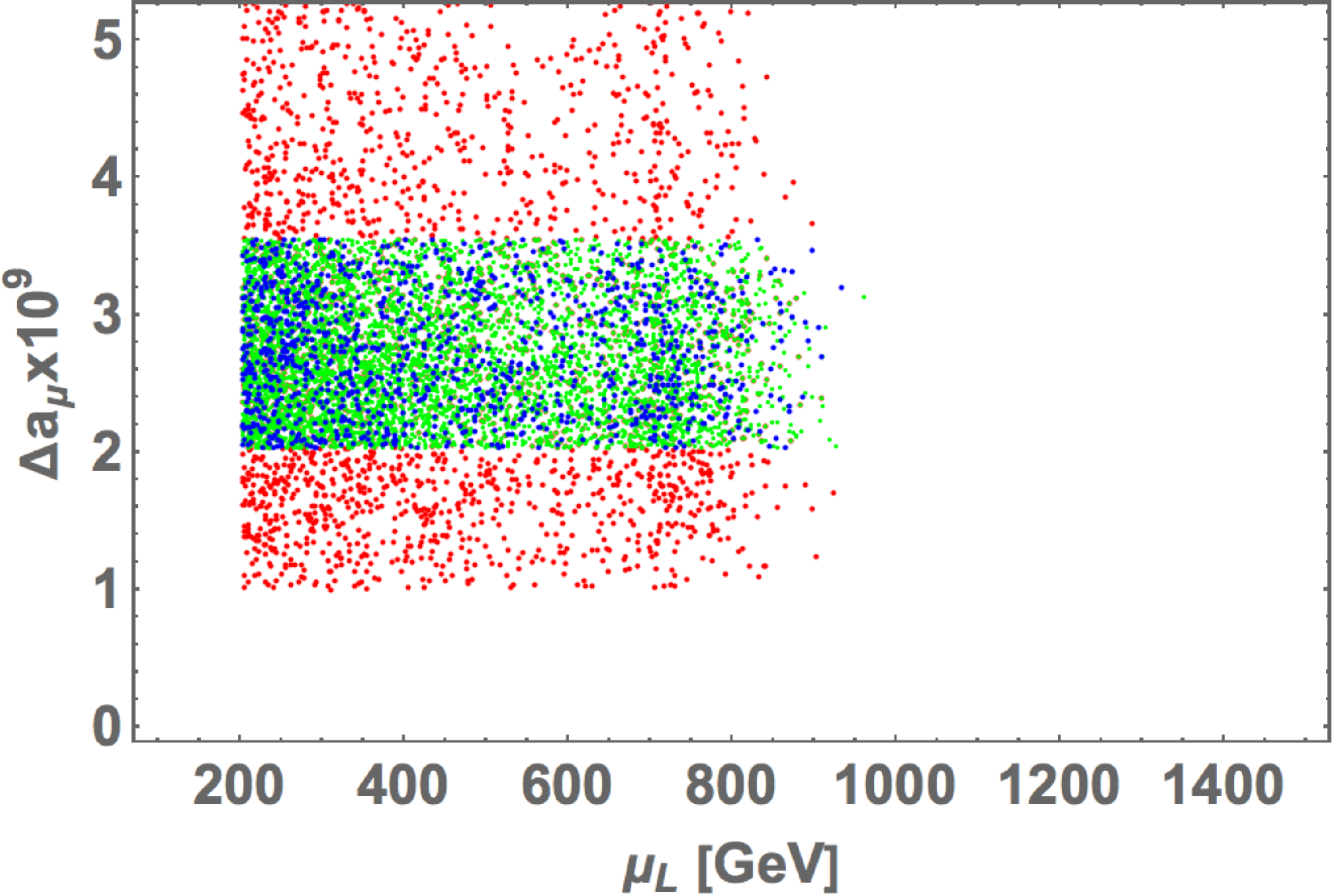}
\caption{Lepton anomalous magnetic moments of electron (left) and muon (right) 
for Model A(top), Model  B(middle) and Model C(bottom) .
Blue points satisfy both Eq.(1) and Eq.(5),  green points satisfy only Eq.(1),
and red points satisfy only Eq.(5).}   
\label{fig:2}\end{center}\end{figure}
%%%%%%%%%%%%%%%%%%%
The results are shown in Figure 1. The constraints for both $g-2$ give the upper bounds of $\mu_L$
as follows: 
\eqn{
\mu_L<375 \mbox{GeV} (\mbox{Model A}), \quad
\mu_L<660 \mbox{GeV} (\mbox{Model B}), \quad
\mu_L<940 \mbox{GeV} (\mbox{Model C}).
}
While the contributions to both $g-2$ from the extra higgsino
are suppressed in Model A,  this contributions are not suppressed and 
raise the upper bound of $\mu_L$ in Model B.
In Model C, an accidental degeneracy of the  diagonal elements of the chargino mass matrix
enhances the off-diagonal elements of the mixing matrices $U,D$, which 
raises the upper bound of $\mu_L$ further.
Eq.(7) and Eq.(206) give the condition:
\eqn{
\frac{Y^{EI}_1}{Y^{EI}_2}\simeq -7,
}
which is satisfied for the numerical calculation as shown in Figure 2.
There is a tendency that
the condition $Y^{EI}_1/Y^{EI}_2>-4$ (green points) gives the smaller $|\Delta a_e|$ and
the condition $Y^{EI}_1/Y^{EI}_2<-10$ (red points) gives the smaller $|\Delta a_\mu|$.
The allowed  region in $\Delta a_e-\Delta a_\mu$ plane is dominated by the
blue points which satisfy the condition $-10\leq Y^{EI}_1/Y^{EI}_2\leq -4$.

\begin{center}
\begin{figure}
\includegraphics[width=7cm,pagebox=cropbox,clip]{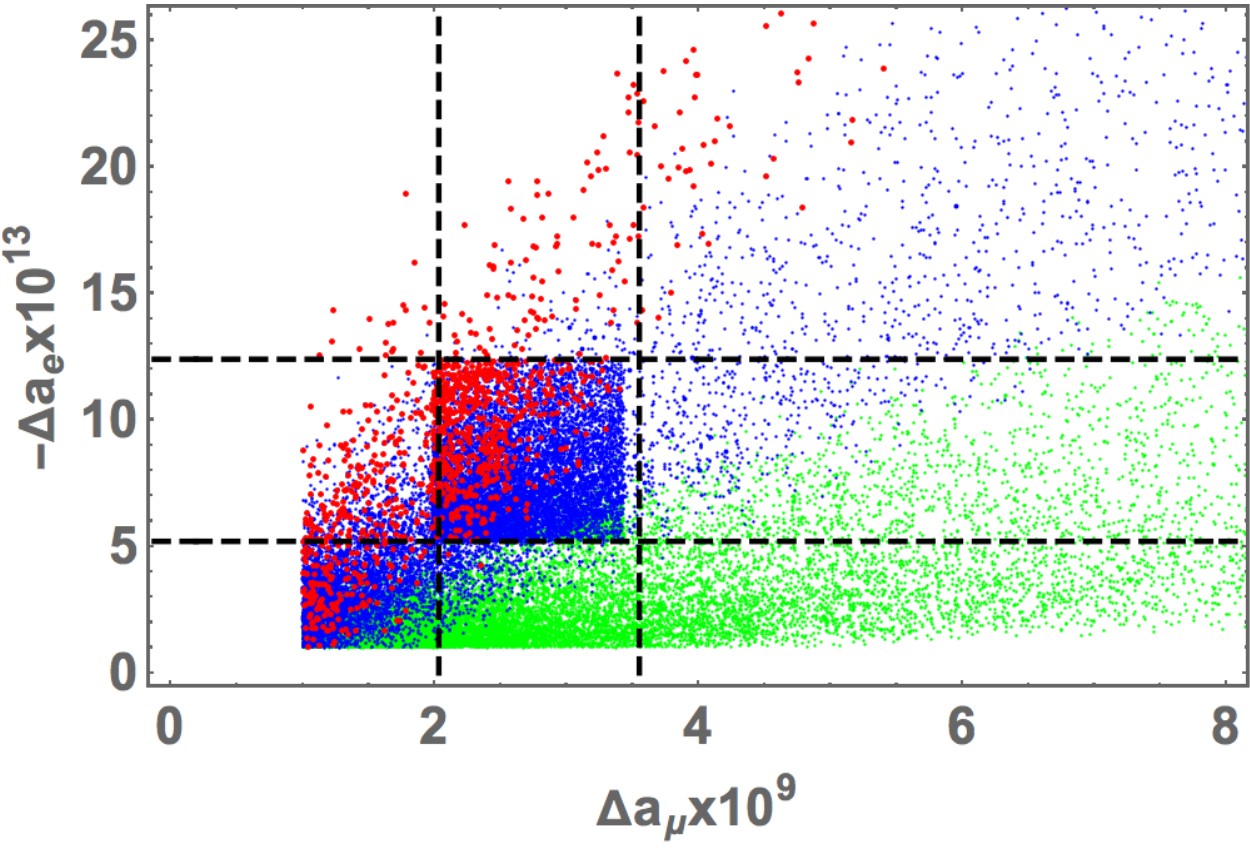}
\caption{Anomalous magnetic moments of electron and muon for Model B.
Green,  blue and red points correspond to
$Y^{EI}_1/Y^{EI}_2>-4$, $-10\leq Y^{EI}_1/Y^{EI}_2\leq -4$ and 
$Y^{EI}_1/Y^{EI}_2<-10$ respectively. 
The vertical (horizontal) dotted lines mean the $1\sigma$ bounds for the muon (electron)
anomalous magnetic moment. }
\end{figure}
\end{center}

%%%%%%%%%%%%%%%%%%%%%%%%%%%%%%%%%%%%%%%%%%%%%%%
% sec7
%%%%%%%%%%%%%%%%%%%%%%%%%%%%%%%%%%%%%%%%%%%%%%%

\section{Conclusions}

We have considered an $S_4$ flavor symmetric  extra U(1) model 
that accounts for dark matter and a baryon asymmetry.
In this model, we assume that dark matter is dominated by an axion
so that the smallness of the up quark mass is understood by the smallness of the Peccei-Quinn scale.
Furthermore, we assume that the muon mass is induced by the result of
the symmetry breaking of the $S_3$ subgroup.
In this case, successful resonant leptogenesis requires TeV scale RHNs,
therefore it may be possible to verify the nature of the RHN by a future collider.
As the TeV scale seesaw mechanism requires very small neutrino Yukawa couplings,
the most relevant interaction of the RHN for a collider experiment is the interaction with the G-Higgs.
Our model also accounts  for two lepton $g-2$ anomalies  without causing too large  flavor violation.
Our numerical estimation shows that
the typical upper mass bound of the lightest
charged SUSY particle in the loop is about 660 GeV.
We can expect to prove the existence of supersymmetry and the flavor symmetry by 
a future collider.

%%%%%%%%%%%%%%%%%%%%%%%%%%%%%%%%%%%%%%%%%%%%%%%
%%%%%%
\section*{Acknowledgments}
This research was supported by an appointment to the JRG Program at the APCTP through the Science and Technology Promotion Fund and Lottery Fund of the Korean Government. This was also supported by the Korean Local Governments - Gyeongsangbuk-do Province and Pohang City (H.O.). 
H. O. is sincerely grateful for the KIAS member. 
%%%%%%%%%%%%%%%%%%%%%%%%%%%%%%%%%%%%%%%%%%%%%%%

%%%%%%%%%%%%%%%%%%%%%%%%%%%%%%%%%%%%%%%%%%%%%%%
% End of file
%%%%%%%%%%%%%%%%%%%%%%%%%%%%%%%%%%%%%%%%%%%%%%%

\end{document}